\documentclass[a4paper,fleqn]{cas-sc}

\usepackage[authoryear,longnamesfirst]{natbib}

\usepackage{amsmath,amssymb,amsfonts}
\usepackage{algorithmic}
\usepackage{graphicx}
\usepackage{textcomp}
\usepackage{wrapfig}

\usepackage{booktabs} 
\usepackage{subcaption}
\usepackage{url}

\usepackage{enumitem}

\def\tsc#1{\csdef{#1}{\textsc{\lowercase{#1}}\xspace}}
\tsc{WGM}
\tsc{QE}
\tsc{EP}
\tsc{PMS}
\tsc{BEC}
\tsc{DE}

\definecolor{arsenic}{rgb}{0.23, 0.27, 0.29}
\definecolor{charcoal}{rgb}{0.21, 0.27, 0.31}
\definecolor{hanblue}{rgb}{0.27, 0.42, 0.81}
\definecolor{blue-ncs}{rgb}{0.0, 0.53, 0.74}
\definecolor{awesome}{rgb}{1.0, 0.13,0.32}
\definecolor{darkgreen}{rgb}{0, .4,0}
\definecolor{purple}{rgb}{.55, .2,.9}

\begin{document}
	
\let\WriteBookmarks\relax
\def\floatpagepagefraction{1}
\def\textpagefraction{.001}
\shorttitle{DiffPhysCam: Differentiable Physics-Based Camera Simulation for Inverse Rendering and Embodied AI}
\shortauthors{B. Chen et~al.}

\title [mode = title]{DiffPhysCam: Differentiable Physics-Based Camera Simulation for Inverse Rendering and Embodied AI}

\author[1]{Bo-Hsun Chen}[orcid=0009-0001-1913-440X]
\cormark[1]


\affiliation[1]{
	organization={Simulation-Based Engineering Lab, University of Wisconsin-Madison}, 
    city={Madison},
    citysep={}, 
    postcode={53706}, 
    state={WI},
    country={USA}
}

\author[1]{Nevindu M. Batagoda}

\author[1]{Dan Negrut}

\cortext[cor1]{Corresponding author}
\nonumnote{Email address: bchen293@wisc.edu}

\begin{abstract}
Generating synthetic images that closely mimic those from real cameras is instrumental in training visual models and enabling end-to-end visuomotor learning. We introduce \textit{DiffPhysCam}, a differentiable camera simulator designed to support robotics and embodied AI applications by enabling gradient-based optimization in visual perception pipelines. Differentiable rendering also allows inverse reconstruction of real-world scenes as digital twins, facilitating simulation-based robotics training. Existing virtual cameras offer limited control over intrinsic settings, have difficulty capturing optical artifacts, and lack tunable calibration parameters. DiffPhysCam addresses these limitations through a multi-stage pipeline that provides fine-grained control over camera settings, models key optical effects such as defocus blur, and supports calibration with real-world data. It enables both forward rendering for image synthesis and inverse rendering for 3D scene reconstruction, including mesh and material texture optimization. We show that DiffPhysCam enhances robotic perception performance in synthetic image tasks. As an illustrative example, we create a digital twin of a real-world scene using inverse rendering and use it to set up a virtual experiment in a multi-physics simulation, in which we demonstrate navigation of an autonomous ground vehicle using images generated by DiffPhysCam. The code, data, and output results associated with this paper are available online for reproducibility at \cite{link_chen2026diffphyscam-camcaliexp,link_chen2026diffphyscam-novelviewsynthesis,link_chen2026diffphyscam_data}.
\end{abstract}




\maketitle
\section{Introduction}
\label{sec:intro}
\subsection{Motivations}
\label{subsec:motivations}
In artificial general intelligence (AGI) research, simulation plays an important role in developing embodied agents that perceive and act in complex, multi-physics environments. Such agents simultaneously engage in diverse sub-tasks, from visual perception to mechanical interaction, as exemplified by humanoids performing household manipulation or autonomous ground vehicles (AGVs) operating in rugged outdoor settings.

Photorealistic camera sensor simulation is gaining increasing attention in the robotics and AGI communities. Frameworks such as \textit{BlenderProc}~\cite{denninger2023blenderproc2} and integrated solutions within robotic simulators -- like \textit{OmniGibson}~\cite{li2023behavior-1k}, \textit{VISTA}~\cite{amini2022vista}, and \textit{Habitat}~\cite{puig2023habitat} -- enable the generation of high-fidelity synthetic images for perception tasks. These virtual cameras are frequently used to train perception algorithms entirely in simulation, with the learned models then transferred to real-world robotic applications. The fidelity of virtual cameras is critical to the success of data-driven visual perception algorithms and end-to-end visuomotor learning, both of which rely on large volumes of visual training data. However, most existing virtual cameras expose only a limited set of configurable parameters, restricting the ability to simulate key optical effects including motion blur, vignetting, defocus blur, lens distortion, exposure variation, and noises, which reduces photorealism. Moreover, their camera models rarely offer tunable internal model parameters that allow for precise calibration against real-world cameras, hindering simulation-to-reality (sim-to-real) transfer.

To design, train, and test embodied AI agents in simulation before real-world deployment, it is important to ensure that these agents generalize well to physical environments, without suffering from a significant sim-to-real gap. This requires constructing high-fidelity digital twins of real-world scenes, which serve as virtual environments for training and evaluation. Inverse rendering provides an efficient approach for building such digital twins from sensor data, enabling synthetic image generation, large-scale data synthesis, and validation of perception algorithms in simulation. However, state-of-the-art inverse rendering methods, such as 3D Gaussian Splatting \cite{kerbl2023gaussian} and Visual Geometry Grounded Transformer (VGGT) \cite{wang2025vggt}, typically assume idealized pinhole camera models and interpret photographs as direct radiance samples. This assumption overlooks critical optical effects introduced by real camera systems, e.g. defocus blur and exposure variation, thereby limiting both the photorealism and fine-grained user control of synthetic imagery.

This contribution introduces a physics-based camera simulator whose tunable model parameters are calibrated against real-world cameras through physical experiments. Because the simulator is also differentiable, it can be integrated after a differentiable renderer, allowing optical artifacts to be incorporated into inverse rendering pipelines and novel image synthesis.

\subsection{Contributions}
\label{subsec:contributions}
Our main contributions are summarized as follows:
\begin{itemize}[leftmargin=*]
	\item A differentiable physics-based camera simulator, \textit{DiffPhysCam}, is proposed. It provides comprehensive camera setting parameters for users to control and generate diverse optical effects, and tunable camera model parameters for customizing characteristics of the camera simulator.
	
	\item Physical experiments were designed and conducted to calibrate the camera model parameters, ensuring accurate alignment with the real camera.
	
	\item The proposed camera simulator can be concatenated after a differentiable renderer for mesh reconstruction and material texture optimization.
	
	\item We present a newly collected small-scale photo dataset of two real-world scenes, called \textit{MythVillage}. It features natural objects like rocks, wooden structures, and artificial succulent plants, all arranged on the ground to mimic wild fields. This dataset, annotated with detailed camera setting parameters, is publicly available online at \cite{link_chen2026diffphyscam_data}.
	
	\item As an example, we demonstrate a multi-physics simulation of AGV navigation in the digital twin of a real-world scene. The calibrated virtual camera can effectively synthesize new images and reproduce various optical artifacts, including defocus blur, exposure variations, and noises.
\end{itemize}

Note that, although each layer in the proposed camera model is implemented using a typical image processing module, the main contributions of this paper lie in combining and organizing these layers into a unified camera model, innovatively introducing tunable model parameters that characterize the virtual camera, and designing physical experiments for calibrating these parameters to reduce the sim-to-real gap. Together, these contributions provide a calibrated, physics-based camera simulation framework that integrates with differentiable rendering and robotics pipelines. By enabling controllable optical effects and alignment with real cameras, \textit{DiffPhysCam} supports the creation of high-fidelity digital twins that reduce the sim-to-real gap in visual perception and control tasks.

Code, data, and outputs are publicly available for reproducing the results in this paper:  \cite{link_chen2026diffphyscam-camcaliexp} supports Sec.~\ref{subsec:calibrate} and Sec.~\ref{sec:AGV_demo}, and \cite{link_chen2026diffphyscam-novelviewsynthesis} supports Sec.~\ref{sec:exp_steup}.

The remainder of this paper is organized as follows: Sec.~\ref{sec:related_works} reviews related work and state-of-the-art techniques. Sec.~\ref{sec:methods} details the proposed camera model, along with the calibration methodology used to align the simulated camera with its real-world counterpart. Sec.~\ref{sec:exp_steup} describes the integration of our camera simulator with a differentiable renderer, the setups of both simulated and real-world scenes for inverse rendering optimization, and optimization results. Sec.~\ref{sec:AGV_demo} demonstrates the AGV simulation. Finally, Sec.~\ref{sec:conclusion} concludes the paper.

\section{Related Work}
\label{sec:related_works}
Several studies have explored physically-based rendering (PBR) for photorealistic camera simulation and sim-to-real gap validation \cite{gruyer2012modeling,grapinet2013characterization,hasirlioglu2018model-based,hasirlioglu2020general}. Blasinski et al. developed a framework based on the Image System Engineering Toolbox (ISET) for soft-prototyping image acquisition systems applied on autonomous driving \cite{blasinski2018optimizing}. Using the same ISET package, Liu et al. analyzed the impact of camera model parameter variations and post-signal processing operations on the generalization of convolutional neural networks \cite{liu2020neural}. Lyu et al. constructed a physical Cornell box and conducted extensive optical property evaluations to assess the sim-to-real gap of their PBR ray-tracing camera simulator \cite{lyu2022validation}. Despite evaluating the sim-to-real gap, prior contributions did not address how to calibrate camera models through real-world experiments, nor did they provide tunable model parameters that could be adjusted to mitigate this gap. Moreover, few of these models are suitable for inverse rendering tasks. A further limitation is that ISET requires over an hour to synthesize an image \cite{lyu2022validation}, adversely impacting its use in robotic simulation. In contrast, the proposed camera simulator provides calibration parameters and draws on physical experiments to align the simulated camera with a real-world camera, effectively bridging the sim-to-real gap.

Several contributions included photorealistic virtual cameras. For example, a PBR-based virtual camera was integrated into the OmniGibson robotic simulator for visual perception simulation \cite{li2023behavior-1k}. The work by Carlson et al. is similar to our modular and differentiable camera model, but is based on a generative adversarial network to augment images via sensor domain transformation \cite{carlson2018modeling,carlson2019sensor}. These methods lack full control over the camera settings or do not offer physically interpretable model parameters, preventing users from explicitly setting camera parameters for desired optical artifacts. In contrast, our proposed camera simulator offers more comprehensive and physically-grounded control parameters for simulating diverse optical phenomena.

Several studies have investigated differentiable renderers for novel image synthesis and 3D scene reconstruction \cite{mildenhall2020nerf,tewari2020state,tewari2022advances,zhang2021nerfactor,muller2022instant}. For instance, Neural Reflectance Decomposition (NeRD) generates decomposed spatially-varying bidirectional reflectance distribution function (BRDF) properties, including albedo, roughness, metallicity, and normals \cite{boss2021nerd}. Nerfstudio is a framework accommodating multiple neural radiance field (NeRF) models for 3D scene reconstruction \cite{tancik2023nerfstudio}. NVDiffRecMC supports extracting meshes, BRDF material textures, and illumination maps that are compatible with off-the-shelf shaders \cite{hasselgren2022shape}. Recent state-of-the-art methods such as Gaussian Splatting, SuGaR, and VGGT enable high-quality 3D reconstruction of large-scale real-world scenes in just a few minutes \cite{kerbl2023gaussian,guedon2024sugar,wang2025vggt}. However, these differentiable renderers typically model captured images as radiance fields under the assumption of ideal pinhole cameras. As a result, they fail to account for camera-induced artifacts such as defocus blur and exposure variations during optimization and rendering. In contrast, our proposed differentiable camera simulator can be integrated with existing differentiable renderers to incorporate real-world optical effects into both the optimization process and novel view synthesis.

More recently, several NeRF-based methods explicitly model camera optics within differentiable renderers to capture effects such as defocus and depth of field \cite{ouyang2021neural,wang2022nerfocus,wu2022dof-nerf}. For instance, Pidhorskyi et al. introduced a physically based, depth-of-field-aware rendering approach that differentiates a disk blur kernel with respect to its radius using Green's theorem. This enables efficient gradient-based optimization of aperture size and focal plane distance in inverse rendering tasks \cite{pidhorskyi2022depth}. Camera Settings Editing of NeRF (CaSE-NeRF) incorporates camera parameters, including aperture size, focus distance, exposure time, and color temperature, to recover radiance fields, allowing to render defocus blur and exposure variations \cite{sun2023case-nerf}. LensNeRF introduces an aperture-aware NeRF that modulates aperture size to achieve defocus blur by changing the ray directions based on the thin lens model \cite{kim2024lensnerf}. However, these methods entangle camera model parameters with the neural renderer during optimization, restricting the optimized NeRF from using other camera models when rendering. Although these models allow changing the aperture size and focus distance at inference time, the reconstructed scene representations, which are typically neural networks, cannot be separated from their corresponding camera models and concatenated with other camera models. In contrast, we decouple the camera model from the differentiable renderer and calibrate the model parameters using physical experiments. This disentanglement enables independent optimization of mesh and material textures, enhancing the flexibility and applicability of our approach for inverse rendering and novel image synthesis. Although NeuralLens can be decoupled from the inverse rendering pipeline as in our approach, it optimizes only for lens distortion and vignetting effects, and does not account for defocus blur or exposure variations \cite{xian2023neural}.

\section{Methods}
\label{sec:methods}
\subsection{Differentiable physics-based simulator architecture}
\label{subsec:cam_struct}
\begin{figure*}
	\centering
	\includegraphics[width=1.0\columnwidth]{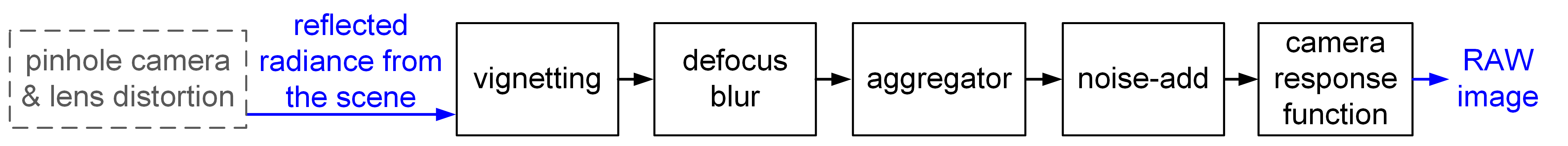}
	\caption{Structure of the proposed differentiable physics-based camera simulator. The blue arrows mark the input and output of the proposed camera simulator.}
	\label{fig:cam_model_struct}
\end{figure*}

\begin{figure}
	\centering
	\includegraphics[width=0.6\columnwidth]{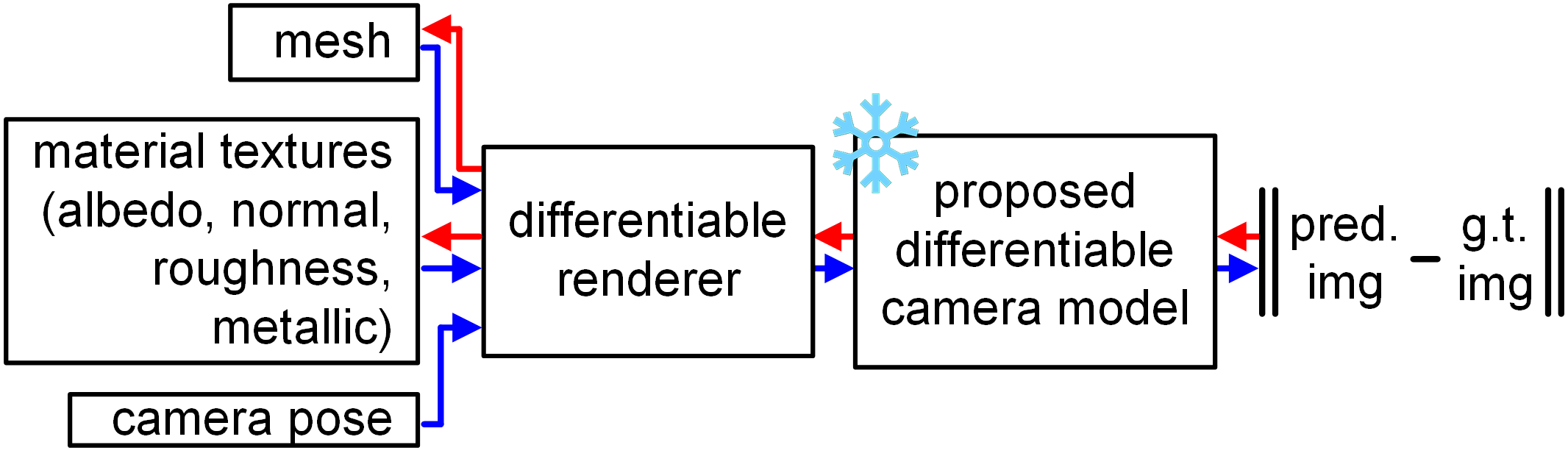}
	\caption{Integration of the camera simulator with a differentiable renderer for mesh and material texture optimization, as well as novel image synthesis. Blue arrows indicate forward inference, while red arrows represent backpropagation for optimizing the mesh and material textures. All camera model parameters are calibrated through physical experiments and then fixed during optimization.}
	\label{fig:concat_NeRF}
\end{figure}

The architecture of the proposed differentiable physics-based camera simulator, DiffPhysCam, is illustrated in Fig.~\ref{fig:cam_model_struct}. DiffPhysCam is implemented as a sequence of modular, differentiable image-processing layers in PyTorch \cite{paszke2019pytorch} that operate on the output of a conventional pinhole camera model, augmenting that scene radiance with optical artifacts. The simulator is used in two modes. In the forward mode, the physically based ray-tracing renderer Chrono::Sensor \cite{asherSensorSimulation2021} models the radiance reflected from the scene, which DiffPhysCam then converts into a realistic synthetic image. In the inverse mode, because every layer is differentiable through PyTorch's auto-differentiation, the simulator can be concatenated after any differentiable renderer (here, we use NVDiffRecMC \cite{hasselgren2022shape}) to optimize the mesh and material textures while accounting for the camera-induced effects; see Fig.~\ref{fig:concat_NeRF}.

The camera simulator operates under several assumptions. First, it adheres to geometric optics principles and disregards wave optics, omitting chromatic aberration and diffraction effects that arise from variations in refractive indices across different wavelengths. Second, the camera simulator uses a perfect hyperbolic lens that conforms to the Gaussian thin lens equation and always remains perpendicular to incident light rays. Consequently, with the exception of radial lens distortion, most spherical aberrations caused by lens thickness, such as coma, astigmatism, and field curvature, are ignored. Furthermore, the camera simulator presumes a single-lens system, so it excludes veiling glare and lens flare that typically result from multi-lens interactions. Each light ray is treated as a composition of red, green, and blue (RGB) components, with their respective intensity mixture determining the color. Lastly, no post-processing operations are applied to the final image output, preserving the raw sensor data.

Each layer in the camera simulator is implemented using PyTorch to enable concurrent pixel-wise computations, facilitating efficient execution on GPUs. The framework provides comprehensive control over key camera setting parameters: aperture number, exposure time, ISO, focus distance, and focal length. Users can change these parameters to simulate common optical effects, such as motion blur, defocus blur, and exposure variations. In addition, each layer incorporates calibration gains, allowing users to calibrate the virtual camera's characteristics to match a real-world camera through physical experiments, thereby reducing the sim-to-real gap. The following paragraphs detail the mathematical models and physical principles underlying each processing layer.

\subsubsection{Pinhole camera and lens distortion}
\label{subsubsec:lens_distortion}
In our pipeline, the camera simulator receives as input the image generated by a pinhole camera model, serving as a baseline representation of scene radiance. In this work, the Chrono::Sensor renderer is used to generate the radiance in simulated scenes. In Chrono::Sensor, the pinhole camera's position serves as the origin of the ray-tracing process, with rays uniformly distributed across the sensor's field of view (FOV), determined by:
\begin{equation}
	\label{eq:fov}
	FOV = 2 \tan^{-1}\left( \frac{w_{sensor}}{2f} \right) \; ,
\end{equation}
where $w_{sensor}$ is the effective sensor width, and $f$ is the focal length. After being emitted, these rays undergo distortion based on the radial lens distortion model, interact with scene geometry, and ultimately reach the light source. The lens distortion parameters are calibrated following the method described in \cite{opencv2024camera}. The output at this stage consists of the irradiance values of the RGB components, measured in $[W/m^2]$.

\subsubsection{Vignetting}
\label{subsubsec:vignetting}
Vignetting is modeled as a reduction in brightness from the center of the image to the periphery and follows a $\cos^4\theta$ attenuation pattern, where $\theta$ represents the ray angle between the optical axis and the ray traveling from the pixel to the aperture center. This attenuation arises from the reduction in projected aperture and pixel areas (both decreasing proportionally to $\cos\theta$) and an inverse-square relationship with the distance from the aperture center to the pixel (contributing a $\cos^2\theta$ factor). The effect is modeled as:
\begin{subequations}
	\begin{equation}
		\label{eq:vignet}
		\left( 1 - \frac{y_{ij}}{x_{ij}} \right)= G_{vignet} \cdot \left( 1 - \cos^4 \left( \theta_{ij} \right) \right) \;,
	\end{equation}
	with the ray angle $\theta_{ij}$ defined as
	\begin{equation}
		\label{eq:ray_angle}
		\theta_{ij} = \tan^{-1} \left( \frac{\sqrt{a_{ij}^2+b_{ij}^2}}{f} \right) \;,
	\end{equation}
\end{subequations}
where $y_{ij}$ and $x_{ij}$ are the output and input pixel intensities, respectively, $(a_{ij}, b_{ij})$ are the pixel coordinates on the image plane, $f$ is the focal length, and $G_{vignet}$ is the vignetting falloff gain. The output remains in irradiance units.

\subsubsection{Defocus blur}
\label{subsubsec:defocus_blur}
\begin{figure}
	\centering
	\includegraphics[width=0.4\columnwidth]{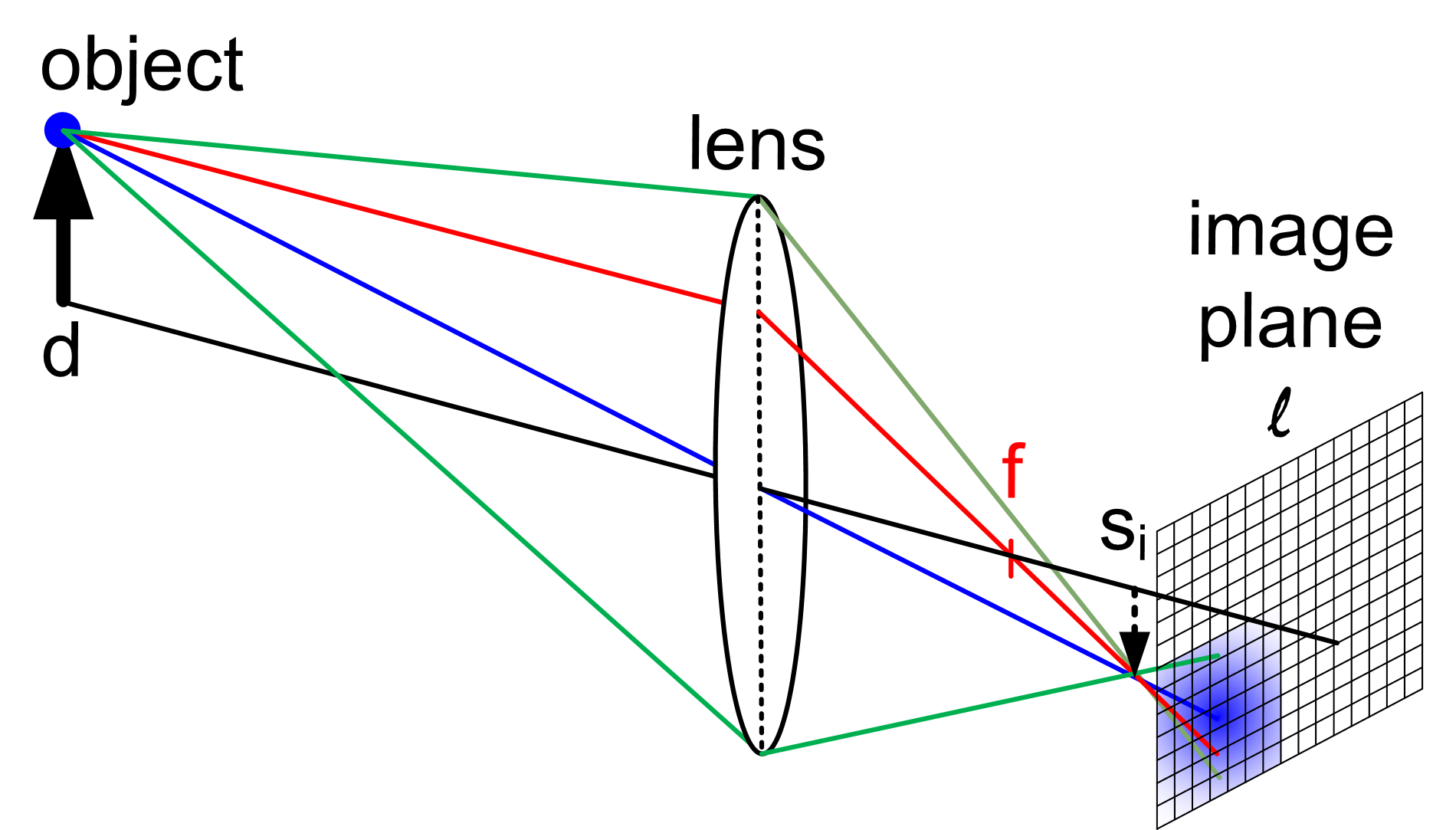}
	\caption{Physical depiction of defocus blur using the Gaussian thin lens model, where $d$ is depth of the scene point, $f$ is focal length, $l$ is the image plane distance, and $U$ is focus distance, and $1/d + 1/s_i = 1/f$ and $1/l + 1/U = 1/f$.}
	\label{fig:defocus_blur_phys}
\end{figure}

Defocus blur is introduced by applying blur kernels to simulate depth of field. According to the Gaussian thin lens model (Fig.~\ref{fig:defocus_blur_phys}), scene points located further away from the focal plane appear as more blurred circles on the image plane. The diameter of these circles of confusion, called defocus blur diameter $D_{ij}$, is computed in units of pixels as:

\begin{subequations}
	\label{eq:defocus_blur}
	\begin{equation}
		\label{eq:defocus_blur_diameter}
		D_{ij} = \frac{G_{defocus} \cdot f^2 \cdot \left| d_{ij} - U \right|}{N \cdot C \cdot d_{ij} \cdot (U - f)} \;,
	\end{equation}
	where $G_{defocus}$ is the defocus gain to be calibrated, $f$ is the focal length, $N$ is the aperture number, $C$ is the pixel size, $U$ is the focus distance, and $d_{ij}$ is the scene depth at pixel $(i, j)$. The resulting blurring effect is achieved by applying a spatially-varying blur kernel to each input pixel and generating the sparse \textbf{defocus blur weight matrix}. Two types of blur kernels are provided as options: Gaussian and Uniform kernels.
	{\small
	\begin{equation}
		\label{eq:defocus_blur_program}
		\begin{aligned}
			& \text{Gaussian:}\quad w_{ijkl} = \frac{1}{2 \pi \sigma^2} \exp \left( -\frac{(k - i)^2 + (l - j)^2}{2 \sigma^2} \right)\;, \\[0.8em]
			& \text{Uniform:}\quad w_{ijkl} = \frac{1}{D_{ij}^2}\;, \\[0.8em]
			& k = i - \frac{D_{ij}}{2}\;\text{to}\;i + \frac{D_{ij}}{2},\;l = j - \frac{D_{ij}}{2}\;\text{to}\;j + \frac{D_{ij}}{2} \; ,
		\end{aligned}
	\end{equation}}
	where $w_{ijkl}$ is the summation weight contributed by the input pixel $x_{ij}$ to the output pixel $y_{kl}$, and $\sigma = D_{ij}/6$ defines the standard deviation of the Gaussian kernel. The output pixel $y_{kl}$ is 
	\begin{equation}
		\label{eq:sparse_mtrx_multiply}
		y_{kl} = \sum \limits_{i=1}^{img\_h} \sum \limits_{j=1}^{img\_w} w_{ijkl} \cdot x_{ij} \; ,
	\end{equation}
\end{subequations}
where $x_{ij}$ is the input pixel, $img\_h$ and $img\_w$ denote the image height and width in pixels, respectively. The weight matrix $[W]_{ijkl}$ is a large sparse matrix of size $(img\_h \cdot img\_w) \times (img\_h \cdot img\_w)$. For efficient computation, the weight calculation is parallelized on the GPU using the Numba library \cite{lam2015numba}. The final output of this layer remains in irradiance units.

\subsubsection{Aggregator}
\label{subsubsec:aggregator}
The aggregator layer accumulates irradiance over time and pixel area to compute the total electron energy per pixel during exposure. This transformation accounts for the scene illumination, sensor characteristics, and camera settings. The aggregated energy is computed as:
\begin{equation}
	\label{eq:aggregator_program}
	y_{ij} = G_{aggregator} \cdot x_{ij} \cdot L_{scene} \cdot \frac{C^2}{N^2} \cdot t \cdot QE_{R/G/B} \;,
\end{equation}
where $y_{ij}$ is the total electron energy collected at pixel $(i, j)$, and $x_{ij}$ is the input irradiance. The gain $G_{aggregator}$ to be calibrated adjusts the overall sensitivity of the virtual camera, $L_{scene}$ corresponds to the illumination intensity of the scene, $C$ represents the pixel size, and $QE_{R/G/B}$ is the quantum efficiency to convert incoming photons into electrons for each RGB color channel. The aperture number $N$ controls the amount of incoming light, and $t$ denotes the exposure time. The output of this layer comes in energy units, Joules, and represents the total charge accumulated in each pixel over the exposure period.

\subsubsection{Noise addition}
\label{subsubsec:noise_add}
To enhance photorealism, the camera simulator incorporates time-dependent (dark current and hot pixels), time-independent (readout and fixed pattern noise (FPN)), and photon shot noises. First, the time-dependent noises, which include dark current and hot pixels, accumulate over time and are added to the output of the previous aggregator layer by:
\begin{subequations}
	\begin{equation}
		\label{eq:dark_current}
		\mu = y_{aggregator} + D_{dark} \cdot t \;,
	\end{equation}
	where $y_{aggregator}$ is the output from the aggregator layer, $D_{dark}$ is the dark current and hot pixel rate, and $t$ is the exposure time. The complete noise model follows the Poisson-Gaussian distribution, defined as:
	\begin{equation}
		\label{eq:noise_model}
		\begin{aligned}
			&Y = L + N \\
			&L \sim Poisson(\mu)  \approx Gaussian(\mu,\; G_{noise}^2 \mu) \\
			&N \sim Gaussian(0,\; \sigma_{read}^2) \;.
		\end{aligned}
	\end{equation}
\end{subequations}
Here, $Y$ denotes the noisy pixel output, modeled as a random variable. Photon shot noise, originally governed by a Poisson distribution, is approximated by a Gaussian distribution with mean $\mu$ and variance $G_{noise}^2 \mu$, where we introduce $G_{noise}$ as a tunable parameter to scale the noise level. This is an acceptable approximation provided that the number of incoming photons is sufficiently large. Readout noise and fixed-pattern noise (FPN) are modeled jointly as a zero-mean Gaussian distribution with standard deviation $\sigma_{read}$, which we also introduce as a gain parameter to be calibrated, capturing the contribution of time-independent noise sources.

\subsubsection{Camera response function (CRF)}
\label{subsubsec:CRF}
Lastly, the camera response function (CRF) converts the analog electronic signals into pixel digital values through amplification and quantization. The amplification gain is governed by the ISO setting, while the camera response curve may follow different functional forms depending on the sensor and imaging pipeline. Below are several commonly used models:
\begin{equation}
	\label{eq:crf}
	\begin{aligned}
		&\text{(Gamma correction)}\quad y_{ij} = (a \cdot ISO \cdot x_{ij})^{\gamma} \\
		&\text{(Sigmoid)}\quad  y_{ij} = \frac{1}{1 + e^{-a \cdot \log_2(ISO \cdot x_{ij}) - b}} \\
		&\text{(Linear)}\quad  y_{ij} = a \cdot ISO \cdot x_{ij} + b \;,
	\end{aligned}
\end{equation}
where $y_{ij}$ is the final pixel digital value, $x_{ij}$ is the input signal, and $a$, $b$, and $\gamma$ are CRF parameters to be calibrated. The ISO-determined gain is applied before conversion. Different CRFs model various camera types. Typically, traditional film cameras follow sigmoid functions, while modern digital cameras use linear functions. The final output is a RAW image in 16-bit-depth format with values in each pixel RGB channel ranging from 0 to 65535, which enables high dynamic range (HDR) imaging.

\subsubsection{Motion blur}
\label{subsubsec:motion_blur}
\begin{figure}
	\centering
	\includegraphics[width=0.55\columnwidth]{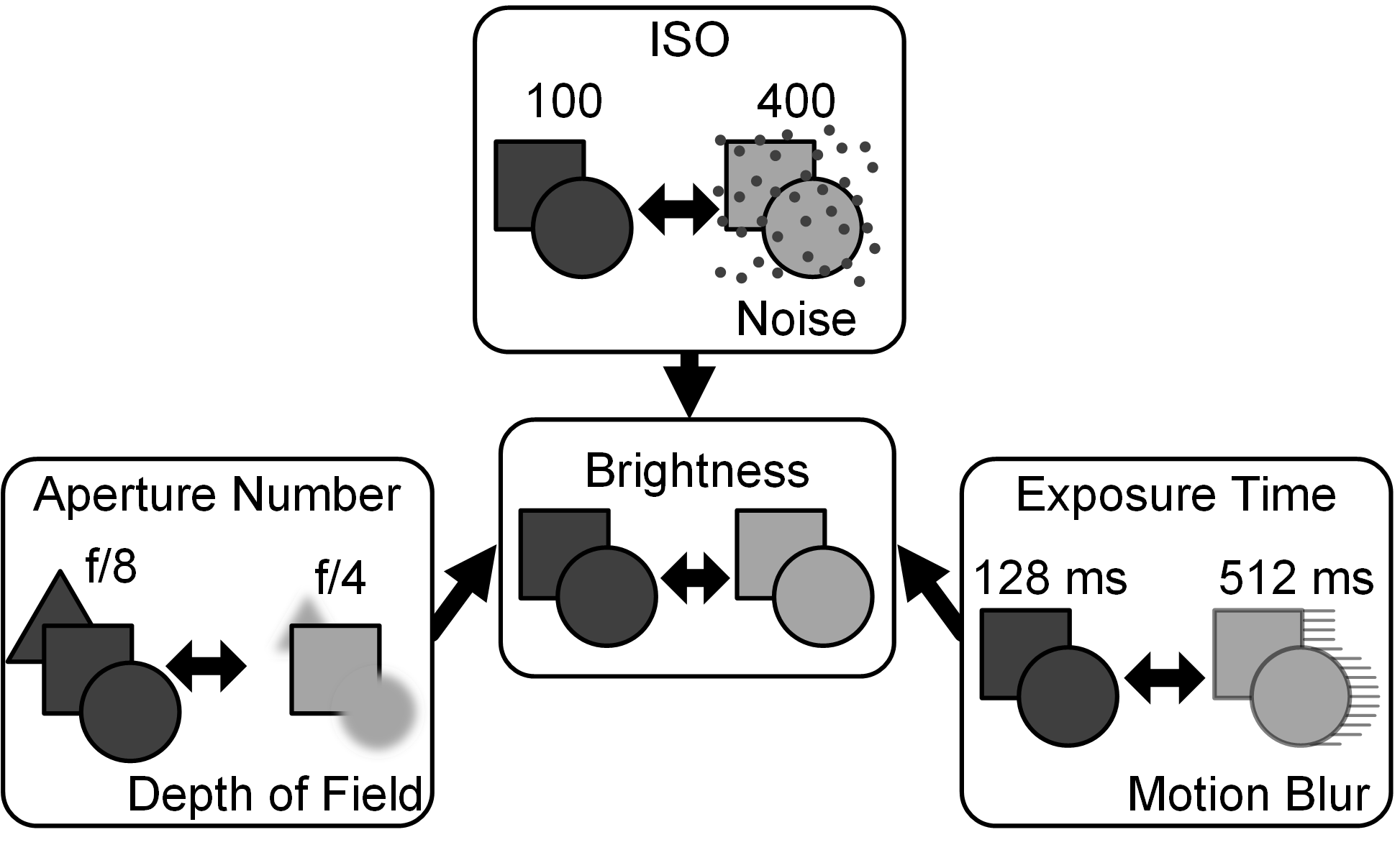}
	\caption{The exposure triangle.}
	\label{fig:expsr_triangle}
\end{figure}

In photography, the exposure triangle describes how three camera setting parameters (aperture number, exposure time, and ISO) jointly control photo brightness, or exposure, with each parameter introducing a distinct visual side effect; see Fig.~\ref{fig:expsr_triangle}. Increasing exposure can be achieved by increasing the exposure time or ISO, or by decreasing the aperture number, which respectively leads to stronger motion blur, stronger noises, or a shallower depth of field. The motion blur induced by the exposure time is calibrated here.

Although motion blur can be superimposed during image post-processing using an optical flow map, OptiX, which underlies Chrono::Sensor, supports higher-fidelity motion blur generation through physical ray-tracing simulation. In real-world photography, the pixel intensity $I$ for a pixel $(u, v)$ is obtained by
\begin{subequations}
	\label{eq:motion_blur}
	\begin{equation}
		\label{eq:px_integral}
		I(u, v) = \int_{t_0} ^{t_0 + t_{expsr}} L(u, v, t)\, dt \approx \sum_{i=1}^{N} L(u, v, T_i)
	\end{equation}
	where $L(u, v, t)$ is the radiance reaching the pixel $(u, v)$ at time $t$, and $t_{expsr}$ is the exposure time. OptiX approximates this integration by accumulating Monte Carlo samples over the exposure time. Motion blur occurs because objects in the scene move relative to the camera during the exposure time. For each sampling ray of a pixel, OptiX randomly selects a time instance within the exposure window $[t_0, t_0 + t_{expsr}]$ and traces the ray as it intersects the scene geometry at that time instance. OptiX then accumulates these ray samples for each pixel to approximate the integration in Eq.~\ref{eq:px_integral}, where $T_i$ is uniformly sampled from $[t_0, t_0 + t_{expsr}]$, and $N$ is the number of samples per pixel.
	
	In this paper, we introduce a model parameter, $G_{motion}$, as a gain multiplied by the exposure time $t_{expsr}$, resulting in an effective exposure window for sampling the time instance $T_i$:
	\begin{equation}
		\label{eq:motion_time_window}
		T_i \sim \textit{Uniform}\left( [t_0, t_0 + G_{motion} \cdot t_{expsr}] \right)\;,
	\end{equation}
\end{subequations}
where the sampled time instance $T_i$ is passed as an argument to the OptiX API function \textit{optixTrace()} to render the motion blur effect.

\subsection{Calibration methods}
\label{subsec:calibrate}
The color camera used in this paper, and for which we calibrated the camera simulator, was the Blackfly\textregistered S BFS-U3-31S4C from Teledyne FLIR, equipped with the ArduCam LN042 5mm prime lens.

\subsubsection{Vignetting falloff gain calibration}
\label{subsubsec:vignet_cali}
\begin{figure}[!t]
	\centering
	\begin{subfigure}[b]{0.35\columnwidth}
		\centering
		\caption{}
		\vspace*{0px}
		\includegraphics[width=\textwidth]{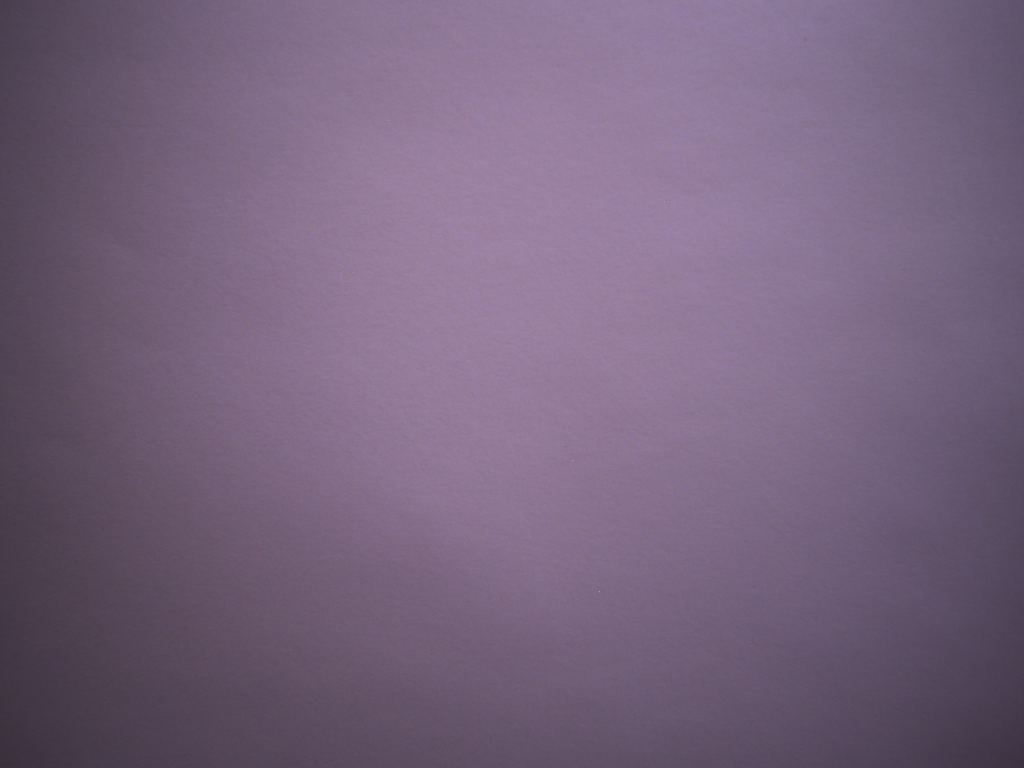}
	\end{subfigure}
	\begin{subfigure}[b]{0.4\columnwidth}
		\centering
		\caption{}
		\vspace*{0px}
		\includegraphics[width=\textwidth]{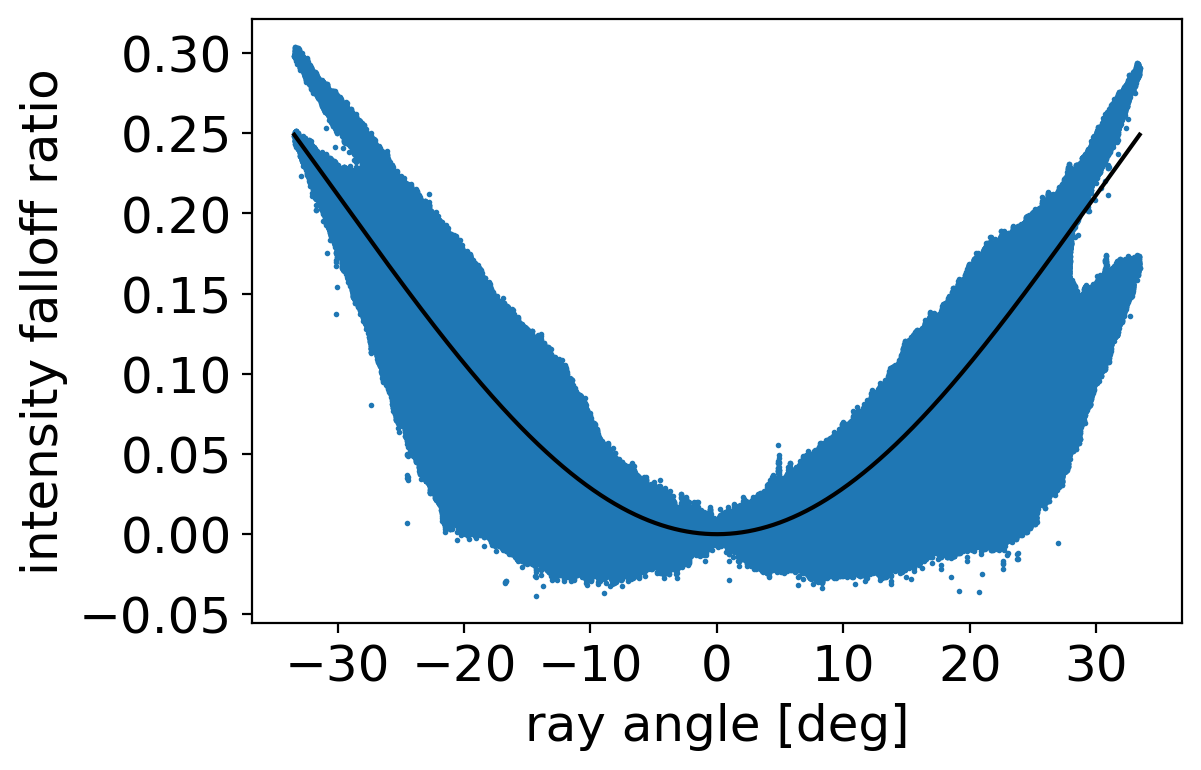}
	\end{subfigure}
	\caption{(a) A photo of a pile of white paper used for vignetting calibration; (b) The relationship between measured intensity falloff ratios and ray angles. Each blue dot represents a pixel, and the black solid line is the regressed intensity falloff function. Negative ray angles correspond to pixels on the left side of the image.}
	\label{fig:vignet_cali_result}
\end{figure}

To calibrate the vignetting effect, photos of a pile of white printing paper were taken in a dark room without direct light sources, relying solely on global illumination. A long exposure time of 1.85 seconds and a wide aperture of F/1.6 were set to ensure sufficient exposure across the entire image. This setting allowed isolation of the vignetting effect, as shown in Fig.~\ref{fig:vignet_cali_result}(a). A total of 20 photos of the same scene were captured and averaged to reduce noises. For each pixel, the corresponding ray angle was computed using Eq.~\ref{eq:ray_angle}. Then, according to Eq.~\ref{eq:vignet}, the vignetting falloff gain $G_{vignet}$ was determined through zero-intercept linear regression, which relates the intensity falloff $(1 - y_{ij}/x_{ij})$ to the ray angle function $(1 - \cos^4(\theta_{ij}))$.

The regression results are shown in Fig.~\ref{fig:vignet_cali_result}(b). The regression yielded an $R^2$ value of 0.603 and a root mean squared error of 0.049. While noticeable variation exists in the measured data, likely due to additional unmodeled factors such as aperture shape and lens structure, the regressed line effectively captures the overall trend, providing a reasonable approximation of the vignetting behavior, despite the real intensity falloff not perfectly aligning with the theoretical prediction.

\subsubsection{Defocus blur gain calibration}
\label{subsubsec:defocus_cali}
\begin{figure}[!t]
	\centering
	\begin{subfigure}[b]{0.243\columnwidth}
		\centering
		\caption{}
		\vspace*{0px}
		\includegraphics[width=\textwidth]{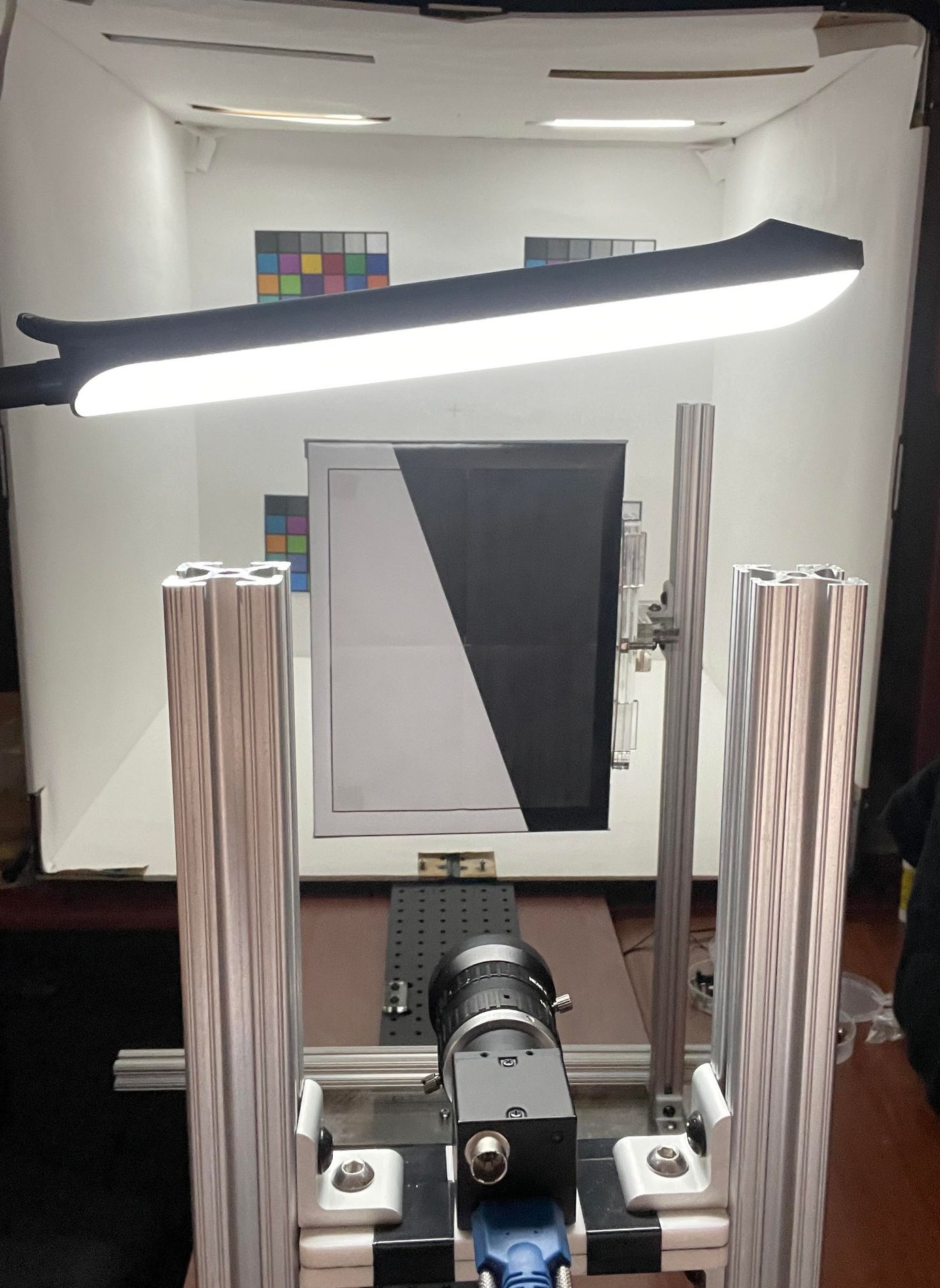}
	\end{subfigure}
	\hfill
	\begin{subfigure}[b]{0.40\columnwidth}
		\centering
		\caption{}
		\vspace*{0px}
		\includegraphics[width=\textwidth]{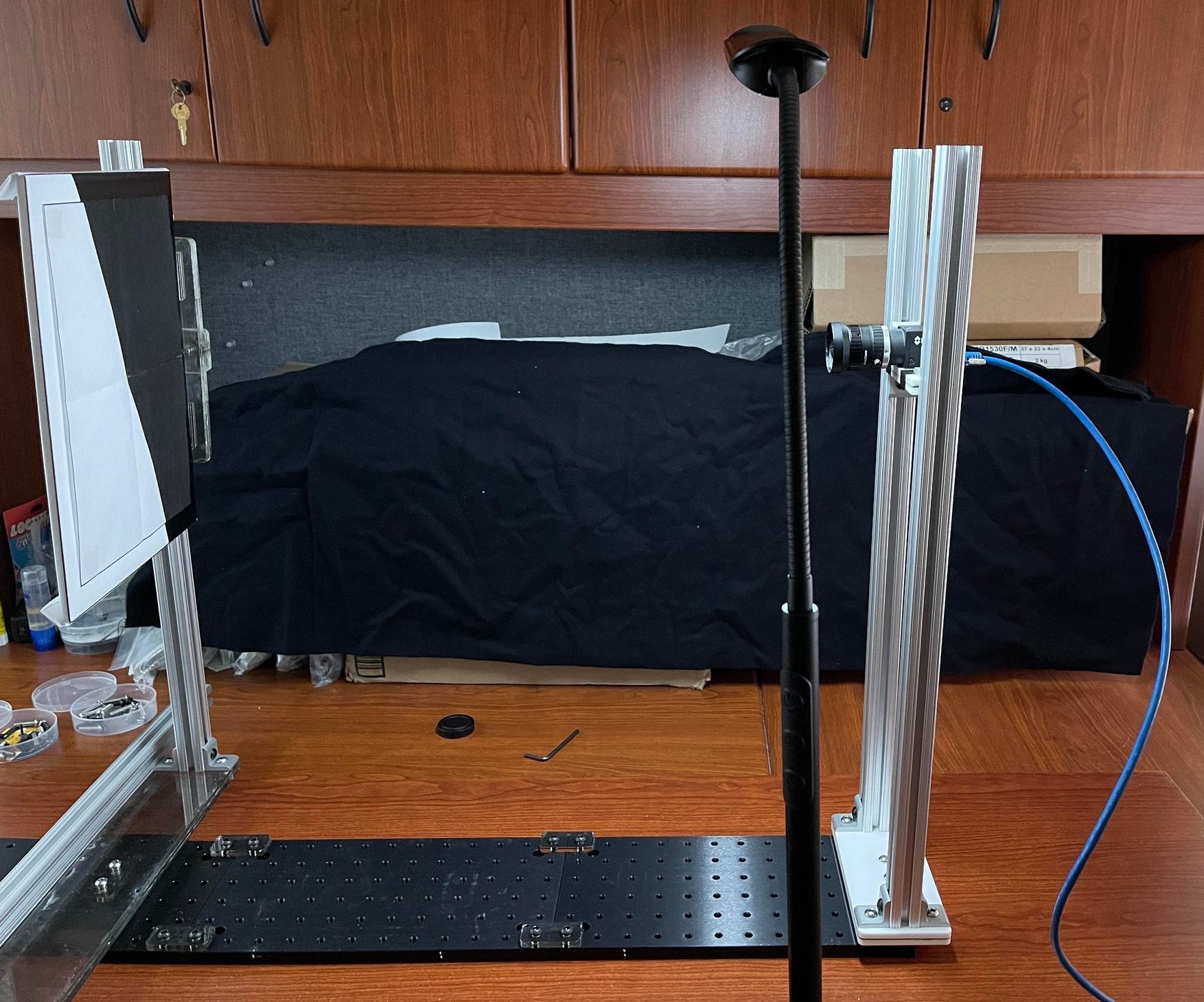}
	\end{subfigure}
	\hfill
	\begin{subfigure}[b]{0.25\columnwidth}
		\centering
		\caption{}
		\vspace*{0px}
		\includegraphics[width=\textwidth]{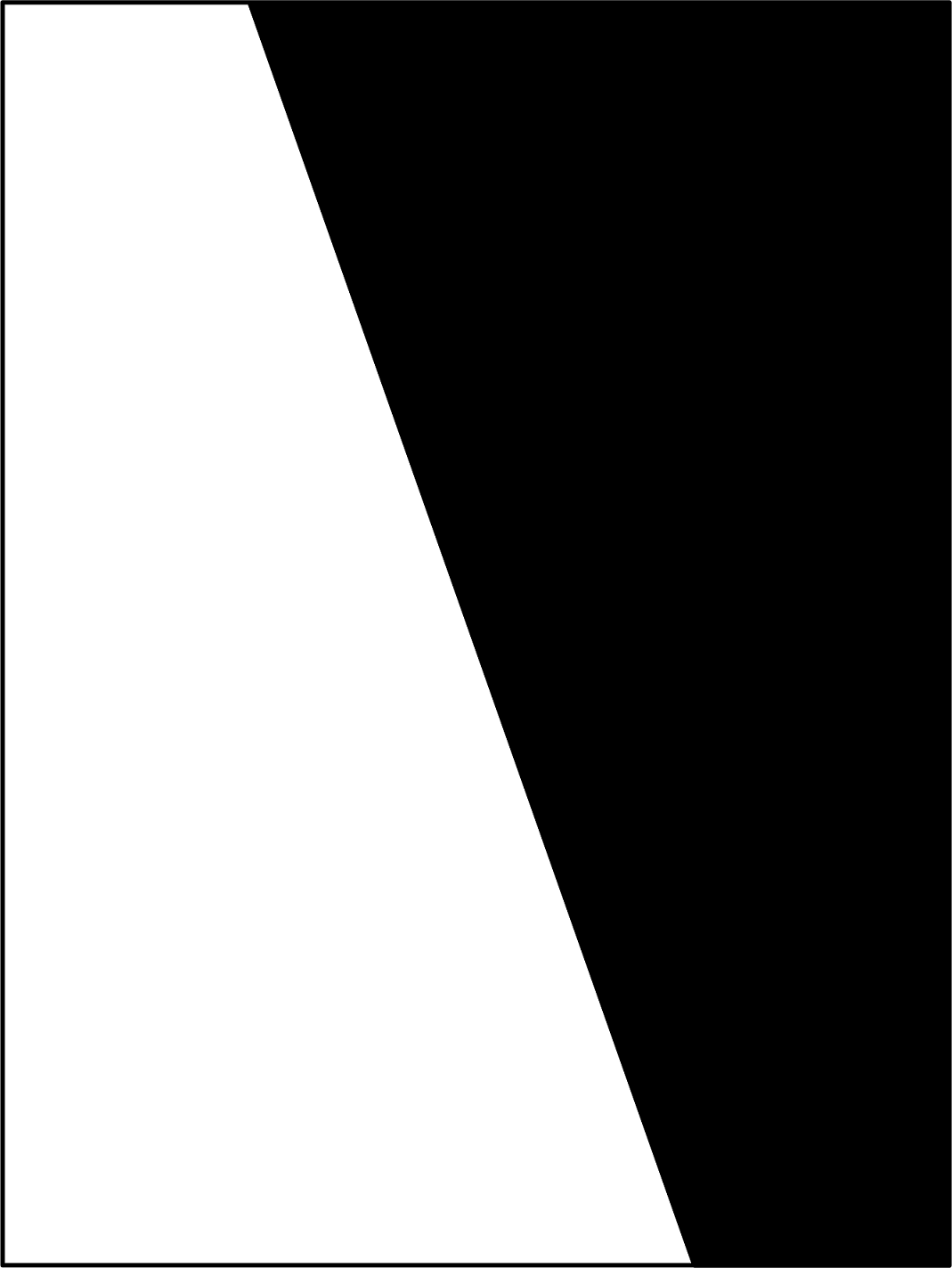}
	\end{subfigure}
	\caption{(a) Back and (b) side views of the experiment setup, and (c) the high-contrast slant pattern used for defocus blur calibration.}
	\label{fig:defocus_setup}
\end{figure}

The experimental setup for defocus blur calibration is illustrated in Fig.~\ref{fig:defocus_setup}. Inspired by the validation approach in \cite{lyu2022validation}, a high-contrast black-and-white pattern with an oblique diagonal division was used to generate and assess defocus blur. According to Eq.~\ref{eq:defocus_blur_diameter}, the defocus blur diameter depends on the scene depth $d$ and the focus distance $U$. In this experiment, the focal length was fixed at 5 mm, the aperture number was set to 1.6 (ensuring the shallowest depth of field), and the pixel size was set according to the camera's documentation. The depth and focus distance were set to all combinations of five discrete values: 162, 312, 612, 998, and 2023 mm, resulting in 25 different experimental conditions.

To quantify the blur, the ``dropping length'' metric was defined, which is analogous to the rise time in control theory. It represents the spatial distance over which pixel intensities transition from 90\% to 10\% of the maximum value. The steps to calculate the dropping length are as follows. First, the intensity of each row in Fig.~\ref{fig:defocus_setup}(c) was normalized to obtain step responses. These step responses were then shifted along the X-axis to align the 50\%-level to the center. Next, the step responses were averaged across all rows, and the distance in millimeters between the 90\% and 10\% levels was defined as the \textbf{dropping length}, quantifying the extent of defocus blur.

To calibrate the defocus blur gain $G_{defocus}$ in Eq.~\ref{eq:defocus_blur_diameter}, an iterative zero-intercept linear regression was applied. A digital twin of the experimental setup was first created using Chrono::Sensor, and a pinhole camera model was used to generate simulated radiance. The camera simulator, initialized with $G_{defocus} = 1.0$, was then applied to synthesize images from the simulated radiance. The synthetic dropping length was extracted from the synthetic images using the same procedure as in the real experiments. For each condition, the ratio between the real and synthetic dropping lengths was computed, and the average of all 25 ratios was used as a scaling factor to update $G_{defocus}$. This iterative process was repeated twice, after which $G_{defocus}$ converged to 0.8 for the Gaussian kernel and to 0.4 for the Uniform kernel, respectively.

\begin{figure*}[!t]
	\centering
	\includegraphics[width=1.0\columnwidth]{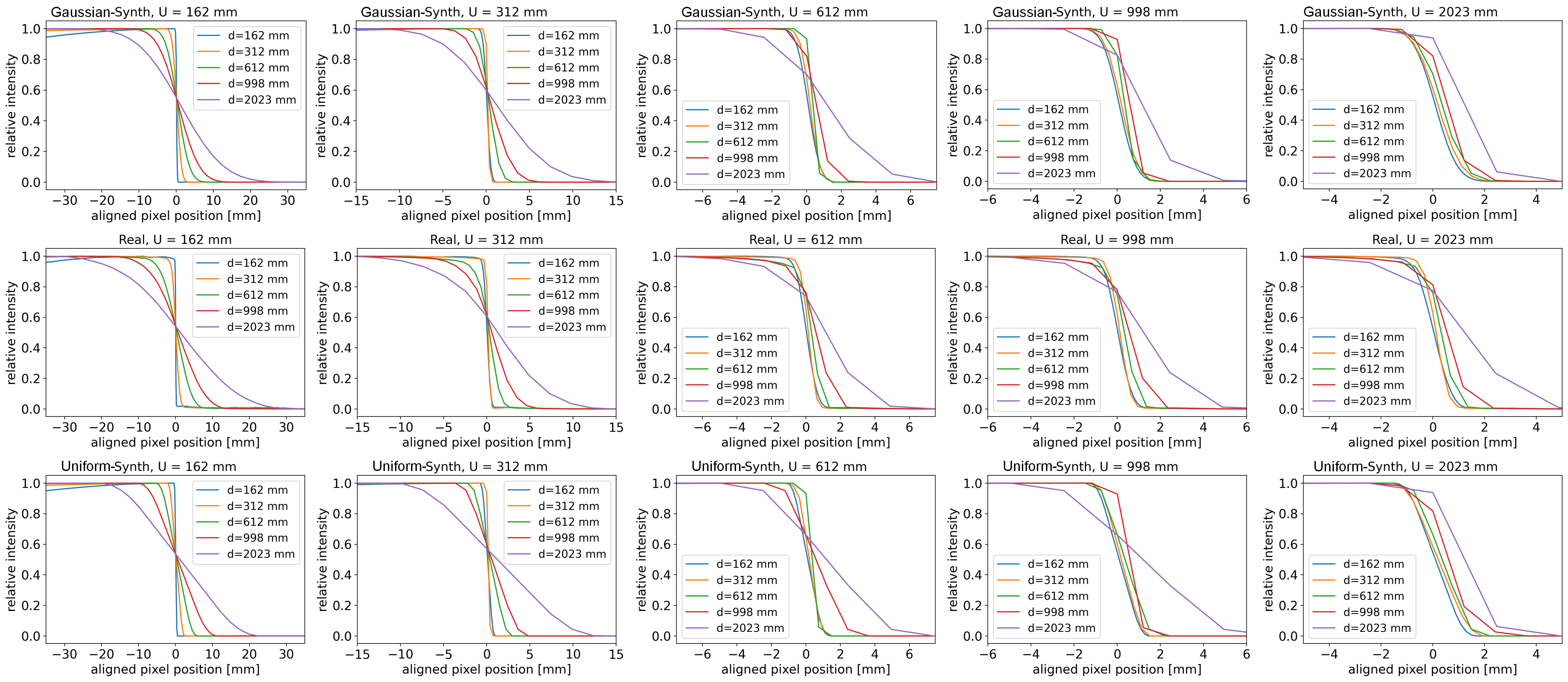}
	\caption{Comparison of step responses among Gaussian-defocused synthesis (top row), reality (middle row), and Uniform-defocused synthesis (bottom row) under different combinations of focus distance $U$ and depth $d$.}
	\label{fig:defocus_steps}
\end{figure*}

\begin{figure*}
	\centering
	\includegraphics[width=1.0\columnwidth]{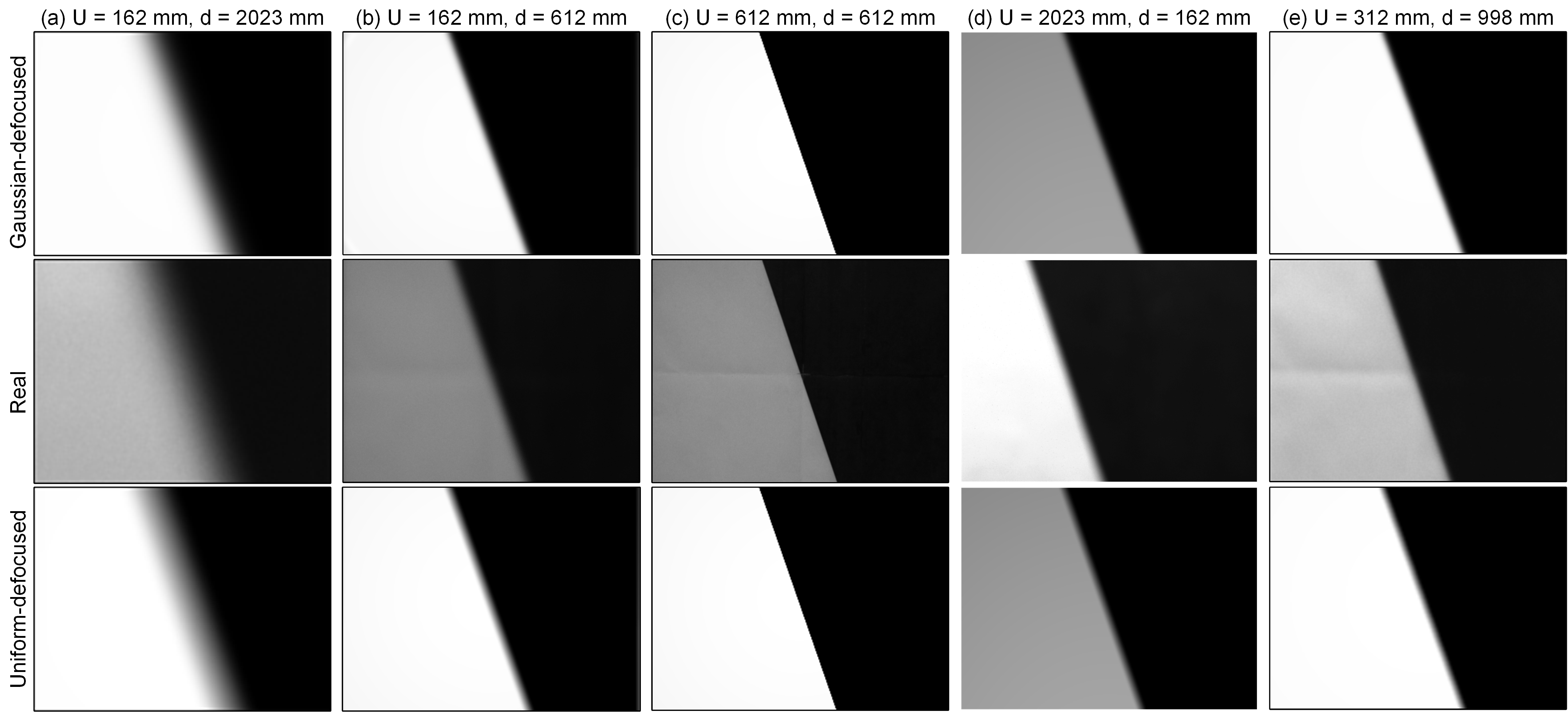}
	\caption{Comparison of cropped defocus-blur patterns among real photos (middle), Gaussian-defocused synthetic images (top), and Uniform-defocused synthetic images (bottom) under different focus distances $U$ and pattern depths $d$ in the scene.}
	\label{fig:defocus_real_synth_cmpr}
\end{figure*}

The final step responses for all real, Gaussian-defocused synthetic, and Uniform-defocused synthetic conditions after calibration are qualitatively shown in Fig.~\ref{fig:defocus_steps}, and quantitative comparisons of the dropping lengths are presented in Table~\ref{table:defocus_drop_lengths}, together with the corresponding error rates. The mean error rates are 18.36\% for the Gaussian-defocused dropping lengths and 24.86\% for the Uniform-defocused dropping lengths, indicating that the Gaussian kernel better fits the experimental results. The results in Table~\ref{table:defocus_drop_lengths} show that the magnitude and range of the synthetic dropping lengths closely match their real counterparts. Moreover, the similarity in trends and patterns between the synthetic and real curves in Fig.~\ref{fig:defocus_steps}, along with the correct ordering of dropping lengths, indicates that the simulated defocus blur closely replicates real-world behavior. Comparisons of the cropped defocused pattern among real photos, Gaussian-defocused synthetic, and Uniform-defocused synthetic images are shown in Fig.~\ref{fig:defocus_real_synth_cmpr} for artifact demonstration.

\begin{table}[!t]
	\centering
	\caption{Dropping lengths (defocus-blur amount) of real photos, Gaussian-defocused, and Uniform-defocused synthetic images after calibration.}
	\renewcommand{\arraystretch}{1.10}
	\begin{tabular}{|c|c|c|c|c|c|}
		\hline
		\multicolumn{6}{|c|}{Real}  \\
		\hline
		$d$ \textbackslash{} $U$ & 162 & 312 & 612 & 998 & 2023 \\
		\hline
		162 & --		& 0.860		& 1.229	& 1.229	& 1.474 \\
		\hline
		312 & 1.912		& --		& 0.956	& 0.956	& 0.956 \\
		\hline
		612 & 8.837		& 3.399		& --	& 1.360	& 1.360 \\
		\hline
		998 & 15.234	& 5.859		& 2.344	& --	& 2.344 \\
		\hline
		2023 & 31.621	& 14.594	& 4.865	& 4.865	& --	\\
		\hline
		\multicolumn{6}{|c|}{Gaussian-defocused Synthetic}  \\
		\hline
		$d$ \textbackslash{} $U$ & 162 & 312 & 612 & 998 & 2023 \\
		\hline
		162			& --		& 0.734		& 1.467		& 1.467		& 1.834			\\
		(err rate)	& --		& (14.7\%)	& (19.4\%)	& (19.4\%)	& (24.4\%)		\\
		\hline
		312			& 2.271		& --		& 1.135		& 1.514		& 1.514			\\
		(err rate)	& (18.8\%)	& --		& (18.7\%)	& (58\%)		& (58\%)	\\
		\hline
		612			& 5.931		& 2.965		& --		& 1.483		& 1.483			\\
		(err rate)	& (32.9\%) 	& (12.8\%) 	& --		& (9.0\%)	& (9.0\%)		\\
		\hline
		998			& 12.097	& 4.839		& 2.419		& --		& 2.419			\\
		(err rate)	& (20.6\%)	& (17.4\%)	& (3.2\%)	& --		& (3.2\%)  		\\
		\hline
		2023		& 24.725	& 14.835	& 4.945		& 4.945		& --			\\
		(err rate)	& (21.8\%)	& (1.7\%)	& (1.6\%)	& (1.6\%)	& --			\\
		\hline
		\multicolumn{6}{|c|}{Uniform-defocused Synthetic}  \\
		\hline
		$d$ \textbackslash{} $U$ & 162 & 312 & 612 & 998 & 2023 \\
		\hline
		162			& --     	& 1.100		& 1.467		& 1.651		& 1.834			\\
		(err rate)	& --		& (27.9\%)	& (19.4\%)	& (34.3\%)	& (24.4\%)		\\
		\hline
		312			& 3.028		& --		& 1.135		& 1.514		& 2.271			\\
		(err rate)	& (58.4\%)	& --		& (18.7\%)	& (58.4\%)	& (137.6\%)		\\
		\hline
		612			& 7.414		& 2.965		& --		& 1.483		& 1.483			\\
		(err rate)	& (16.1\%)	& (12.8\%)	& --		& (9\%)		& (9\%)			\\
		\hline
		998			& 12.097	& 4.839		& 2.419		& --    	& 2.419			\\
		(err rate)	& (20.6\%)	& (17.4\%)	& (3.2\%)	& --		& (3.2\%)		\\
		\hline
		2023 		& 24.725	& 14.835	& 4.945		& 4.945		& --			\\
		(err rate)	& (21.8\%)	& (1.7\%)	& (1.6\%)	& (1.6\%)	& --			\\
		\hline
		\multicolumn{6}{l}{unit: [mm], $d$: scene depth, $U$: focus distance, (.): error rate calculated} \\
		\multicolumn{6}{l}{ relative to the real dropping lengths.} \\
	\end{tabular}
	\label{table:defocus_drop_lengths}
\end{table}

\subsubsection{Exposure gain calibration}
\label{subsubsec:exposure_cali}
Exposure calibration determines multiple parameters across different layers in the camera simulator. These include the product of aggregator gain and quantum efficiency $G_{aggregator} \cdot QE_{R/G/B}$ in the aggregator layer, the dark current $D_{dark}$ in the noise layer, and the bias terms $b_{R/G/B}$ in the CRF layer. By assuming a linear CRF in Eq.~\ref{eq:crf} with the gain parameter $a$ equal to 1, the effect of camera settings on exposure can be expressed as:
{\small
\begin{equation}
	\label{eq:expsr_ctrl}
		DV =G_{aggregator} \cdot QE \cdot \left( \frac{ISO \cdot t \cdot K \cdot R}{N^2} \right)
		+ D_{dark} \cdot ISO \cdot t + b\; ,
\end{equation}}
where $DV$ is the final pixel digital value, $ISO$ is the ISO setting, $t$ is the exposure time, $K$ is a constant product of predefined parameters, $R$ is the input radiance of the camera model, $N$ is the aperture number, $QE$ and $b$ correspond to the quantum efficiency and bias in the CRF for each RGB channel; the remaining parameters have been already explained above.

\begin{figure}[!t]
	\centering
	\includegraphics[width=0.65\columnwidth]{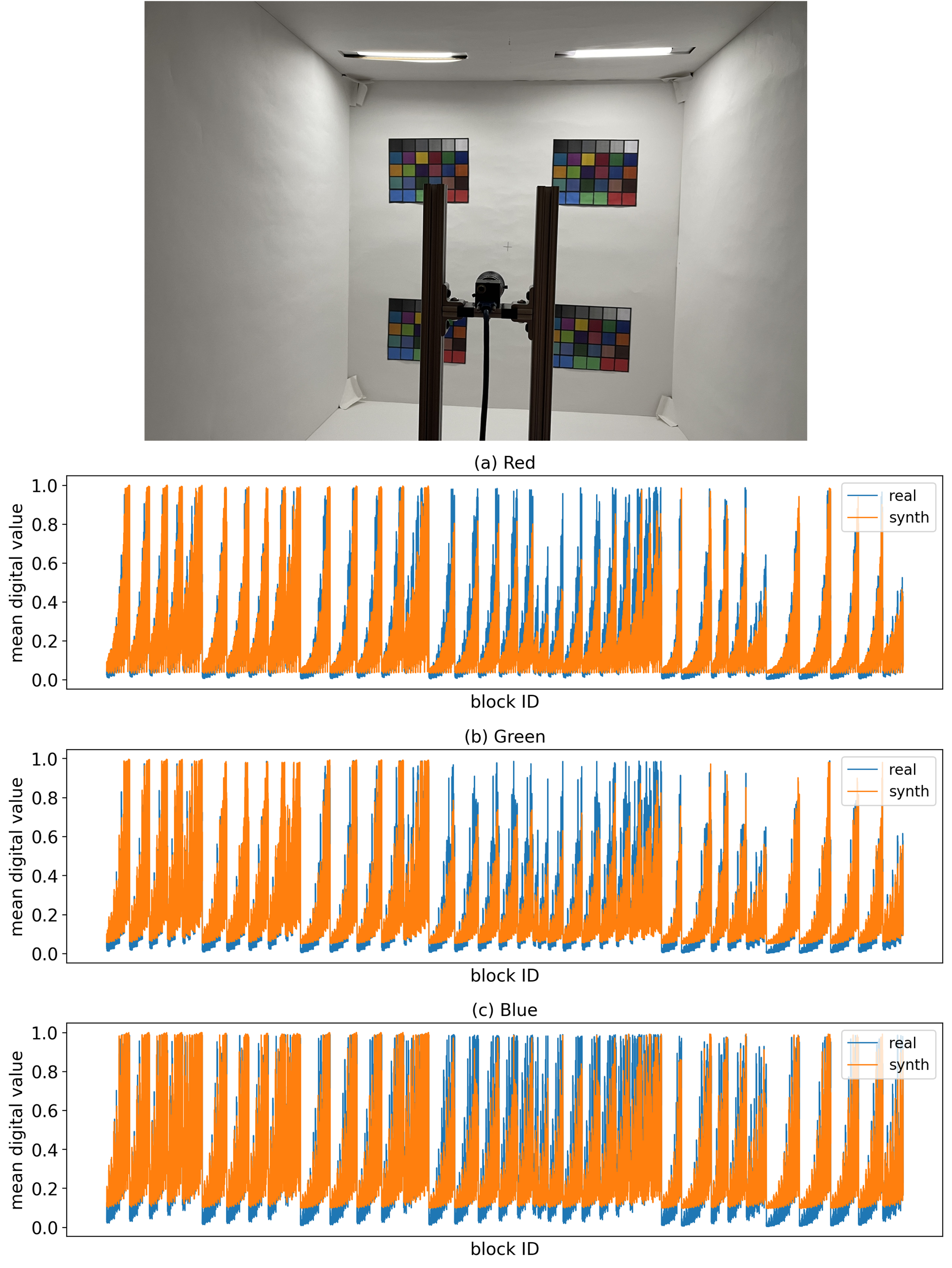}
	\caption{Experimental setup for exposure calibration, and comparison of averaged digital values for each probe between the real photos and calibrated synthetic images for the (a) red, (b) green, and (c) blue channels.}
	\label{fig:expsr_response_cmpr}
\end{figure}

To calibrate these model parameters, a controlled experimental environment was constructed to capture images, as shown in Fig.~\ref{fig:expsr_response_cmpr}. The setup was designed to minimize unknowns and control as many variables as possible. A large Cornell-box-like cube was built to serve as the physical test environment. The cube was constructed from five boards, each with a thickness of 0.25 inches, to block external ambient light. The inner sides of the cube each measured 1 meter, matching the expected operating distance of the camera. Each board was made from high-density fiberboard and covered with white Canson XL Mixed Media (98 lb) drawing paper to provide a uniform reflective surface.

Two 1800-lumen dimmable floor lamps were mounted in 5cm $\times$ 23.5cm rectangular openings on the top board, serving as the light sources. This configuration enabled more uniform illumination across the cube interior compared to using a single circular area light, as is standard in the Cornell box. Both light sources were set to 10\% of their maximum intensity (i.e., 180 lumens), with a color temperature of 6500~K, corresponding to an RGB value of (255, 254, 250) \cite{academo2024colour}.

To evaluate illumination response, four identical Macbeth Color Checkers (MCCs) were attached to the backboard as the target pattern. Each MCC contained an additional row of red, green, and blue color blocks. The MCCs were printed using a high-quality color printer and placed at different locations to assess the effects of different incident light angles. The camera was positioned 95 cm away from the backboard, with its height aligned to the backboard's center and focused at the backboard. A total of 216 exposure combinations of varying exposure times ($t$), aperture numbers ($N$), and ISOs ($ISO$) were used to take the photos. In each photo, 120 colored blocks on MCCs were numbered and set as the probes. The mean R, G, and B channel values of each block were calculated as the probe values. Thus, each photo had 360 probe values to be compared with simulated counterparts.

To construct a simulated digital twin of the experimental setup and generate the corresponding radiance input for the camera simulator, the material properties of the drawing paper needed to be determined first. The Disney BRDF model \cite{burley2012physically} was adopted to represent the material, parameterized by albedo RGB color, specular, roughness, and metallicity. The albedo was assumed to match the \textit{Office Paper} material from the \textit{Physicallybased} database \cite{palmqvist2024physicallybased}. The remaining parameters, i.e., specular, roughness, and metallicity, were estimated using a genetic algorithm, based on measured material data from the \textit{Rough Paper} entry in the \textit{CUReT} database \cite{dana1999reflectance}. All other simulation configurations were kept identical to those used in the physical experimental setup. As a result, each synthetic image contained 360 probe values to match its real-world counterpart.

An iterative multivariate linear regression was performed to calibrate the parameters $(G_{aggregator} \cdot QE,\; D_{dark},\; b)$ in Eq.~\ref{eq:expsr_ctrl}. The parameters were first initialized, and the camera simulator was applied to the simulated radiance under 216 different camera settings, generating a total of 77{,}760 synthetic probe values. Using the synthetic probes as features and the corresponding real probes as labels, multivariate linear regression was then applied to update the parameter values in Eq.~\ref{eq:expsr_ctrl}. This regression process was repeated until convergence (three times). 

The digital values of all color probes in the real photographs and the final calibrated synthetic images are compared in Fig.~\ref{fig:expsr_response_cmpr}. The mean and standard deviation of the absolute errors of digital values in Fig.~\ref{fig:expsr_response_cmpr} are 0.049 $\pm$ 0.056, 0.063 $\pm$ 0.066, and 0.107 $\pm$ 0.107 for the red, green, and blue channels, respectively. The synthetic digital values closely match the real measurements in Fig.~\ref{fig:expsr_response_cmpr}, with small mean errors across all channels, indicating that the exposure calibration regression performs well.

\subsubsection{Noise model calibration}
\label{subsubsec:noise_cali}
\begin{figure}[!t]
	\centering
	\begin{subfigure}[b]{0.4\columnwidth}
		\centering
		\vspace*{0px}
		\includegraphics[width=\textwidth]{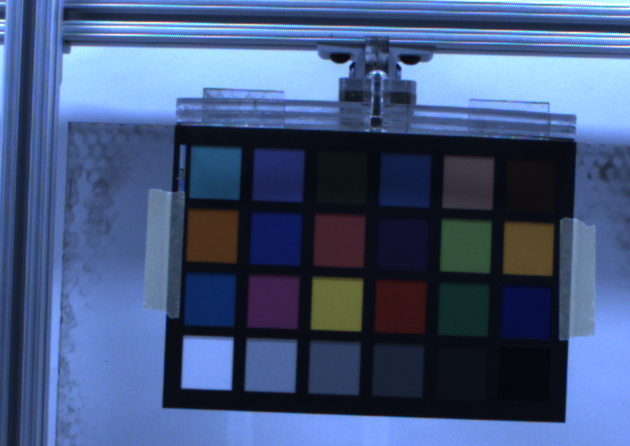}
	\end{subfigure}
	\begin{subfigure}[b]{0.4\columnwidth}
		\centering
		\vspace*{0px}
		\includegraphics[width=\textwidth]{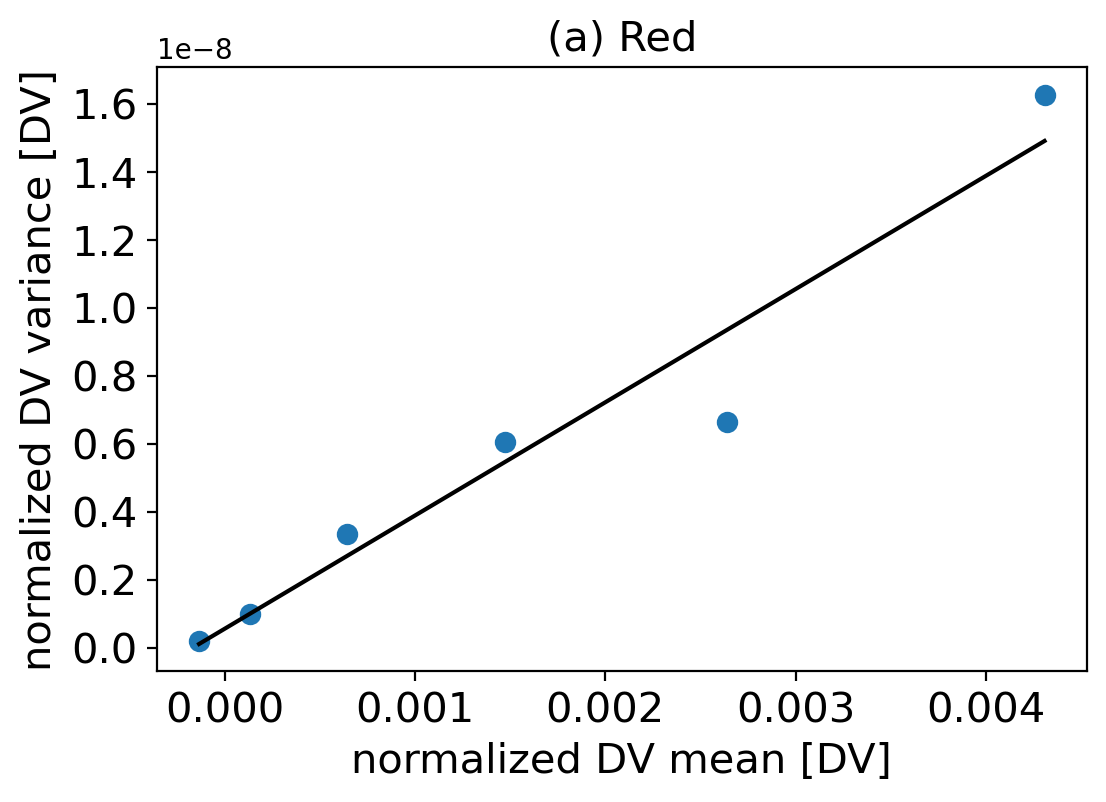}
	\end{subfigure}
	\vspace*{0px}
	\begin{subfigure}[b]{0.4\columnwidth}
		\centering
		\vspace*{0px}
		\includegraphics[width=\textwidth]{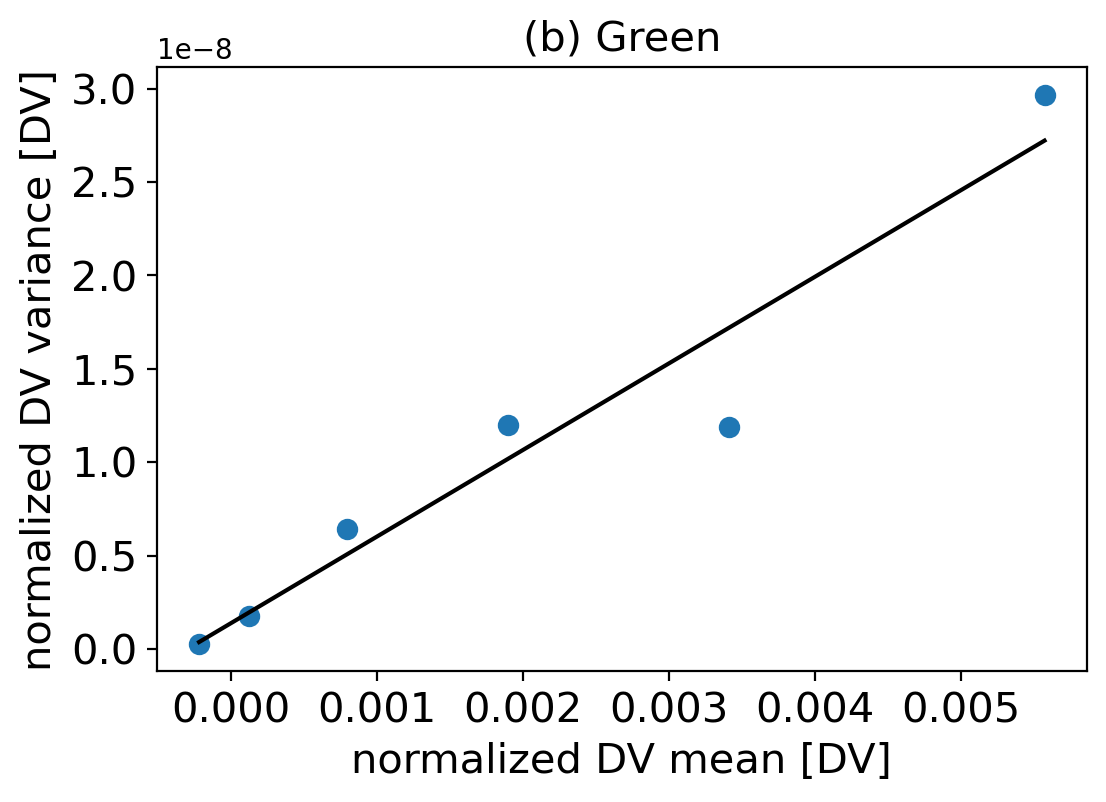}
	\end{subfigure}
	\begin{subfigure}[b]{0.4\columnwidth}
		\centering
		\vspace*{0px}
		\includegraphics[width=\textwidth]{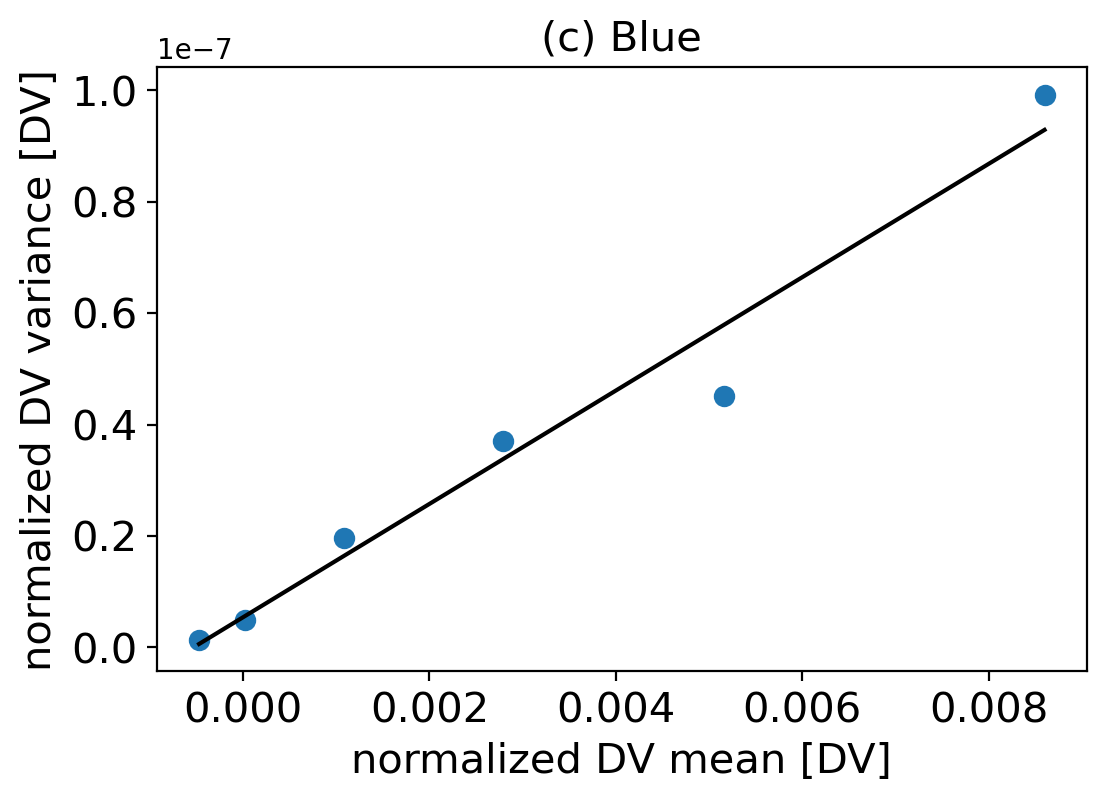}
	\end{subfigure}
	\caption{(a) The photo of MCC for noise calibration; The normalized variances vs. means of measured DVs for the (a) red, (b) green, and (c) blue channels, with the regressed linear functions as shown.}
	\label{fig:noise_cali_result}
\end{figure}

The noise parameters to be calibrated are $G_{noise}$ and $\sigma_{read}$. Assuming a linear camera response function, and following Eq.~\ref{eq:noise_model}, the digital pixel value $I$---also modeled as a random variable---is given by:
\begin{subequations}
	\begin{equation}
		I = ISO \cdot Y + b = ISO \cdot (L + N) + b \; ,
	\end{equation}
	where $b$ is the digital value bias. Then, the expectation of the digital value $\mathbb{E}[I]$ is
	\begin{equation}
		\mathbb{E}[I] = ISO \cdot (\mathbb{E}[L] + \mathbb{E}[N]) + b = ISO \cdot (\mu +0) + b \;,
	\end{equation}
	while the variance of the digital value $\mathbb{V}[I]$ is 
	\begin{equation}
		\begin{aligned}
			\mathbb{V}[I] &= ISO^2 \cdot (\mathbb{V}[L] + \mathbb{V}[N]) \\
			&= ISO^2 \cdot (G_{noise}^2 \cdot \mu + \sigma_{read}^2) \\
			&= ISO^2 \cdot (G_{noise}^2 \cdot \frac{\mathbb{E}[I] - b}{ISO} + \sigma_{read}^2) \\
			&= ISO \cdot G_{noise}^2 \cdot (\mathbb{E}[I] - b) + ISO^2 \cdot \sigma_{read}^2 \;.
		\end{aligned}
	\end{equation}
	Thus, the parameters $G_{noise}$ and $\sigma_{read}$ can be calibrated via linear regression based on:
	\begin{equation}
		\label{eq:noise_cali}
		\left( \frac{\mathbb{V}[I]}{ISO^2} \right) = G_{noise}^2 \cdot \left( \frac{\mathbb{E}[I] - b}{ISO} \right) + \sigma_{read}^2 \;.
	\end{equation}
\end{subequations}
For calibration, a commercial MCC was placed in the customized Cornell box, with two overhead lamps on. Only the last row of six grayscale blocks on the MCC was probed, as shown in Fig.~\ref{fig:noise_cali_result}(a). The camera was set with F/4.0 aperture, ISO 100, and 256 ms exposure time. The mean and variance of pixel intensities were computed for each grayscale block, yielding six data points for regression.

The measured values and regression result are shown in Fig.~\ref{fig:noise_cali_result}(b), where the horizontal axis represents the normalized mean DV for each block, $(\mathbb{E}[I] - b)/ISO$, and the vertical axis represents the normalized variance of the DV, $\mathbb{V}[I]/ISO^2$. The noise model calibration results were imprecise and warrant further investigation, as data points from different camera settings did not align along a single trend line, contrary to theoretical expectations. As a result, only one camera setting was used for calibration. The final noise model parameters were then manually adjusted based on visual comparison with the real photographs.

\subsubsection{Gamma calibration in CRF}
\label{subsubsec:CRF_cali}
The gamma value ($\gamma$) in the CRF was determined by analyzing how pixel digital values changed with variations in exposure, which was controlled by ISO, exposure time ($t$), and aperture number ($N$). The relationship between digital values and exposure follows:
\begin{equation}
	\label{eq:log_dv_to_expsr}
	\log_2 (DV) \propto \gamma \cdot \log_2 (t \cdot ISO \cdot N^{-2})\; ,
\end{equation}
where $DV$ is the final digital value in the photo. By adjusting one control factor of the exposure triangle while keeping the other two constant, the gamma value can be determined through the logarithmic relationship between digital values and exposure changes. Experimental results and corresponding plots confirm that $\gamma = 1.0$, indicating that the camera response function mapping exposure to digital values is linear, i.e., $y = ax + b$. Further details are provided in Appendix~\ref{sec:appndx_gamma}.

\subsubsection{Motion blur calibration}
\label{subsubsec:motion_blur_cali}
\begin{figure}[!t]
	\centering
	\includegraphics[width=0.6\columnwidth]{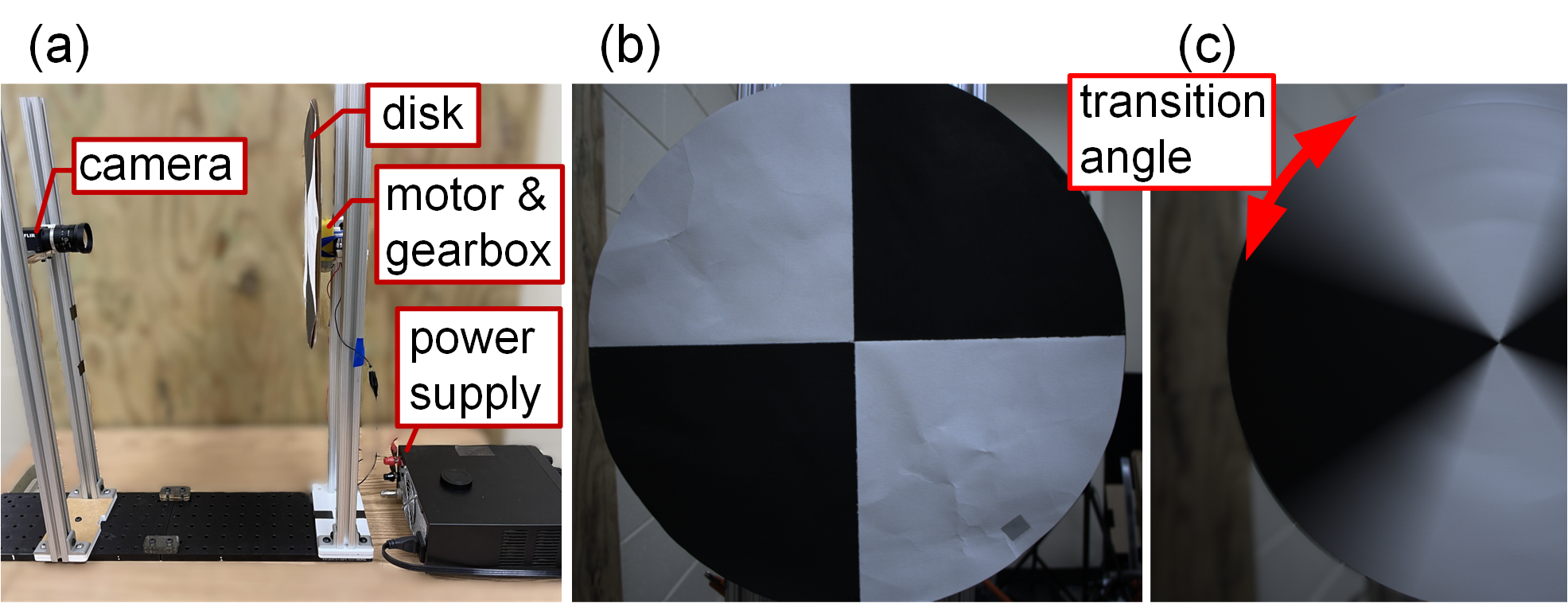}
	\caption{Experimental setup for motion blur calibration: (a) experimental setup, (b) pattern on the rotating disk, and (c) definition of the transition angle, indicated by the dual-headed red arrow.}
	\label{fig:motion_blur_exp}
\end{figure}

The experimental setup for motion blur calibration is shown in Fig.~\ref{fig:motion_blur_exp}(a). A DC motor driven by a power supply was used to generate rotational motion and produce motion blur. A circular disk pattern divided into four alternating black-and-white quadrants was mounted on the DC motor, as shown in Fig.~\ref{fig:motion_blur_exp}(b). When the disk was rotating, motion blur appeared along the boundary between the black and white regions, forming a transitional sector with a color gradient from black to white, as shown in Fig.~\ref{fig:motion_blur_exp}(c). Similar to the ``dropping length'' metric defined for defocus blur calibration in Sec.~\ref{subsubsec:defocus_cali}, the \textbf{transition angle} was used to quantify and calibrate the amount of motion blur. This metric is defined as the spatial angle over which the pixel intensities drop from 90\% to 10\% of the maximum value within the transition sector.

The real-world camera was positioned normal to the plane of the rotating disk to take photos. Six exposure times, 30, 50, 70, 90, 110, and 130 ms, were set to capture different levels of motion blur. Three constant DC voltages, 1.0, 1.5, and 2.0 V, were applied to drive the motor and rotate the disk at different angular speeds. For each driving voltage, the average angular speed of the rotating disk was measured using a digital tachometer. The measured angular speeds, which were used to set up the digital twin of the experiment in simulation, were 3.1259, 4.9480, and 6.9953 rad/sec under 1.0, 1.5, and 2.0 V, respectively. In the following discussion, these three rotation speeds are referred to as low, medium, and high speeds, respectively. In total, 18 experimental conditions were tested, and five photos were captured for each condition. The light source was the ceiling light, which was configured to avoid overexposure under the longest exposure time. The focal length was 5 mm, the aperture number was 2.38, and the ISO was 100.

The calibration procedure was similar to that used for defocus blur. To calibrate the motion-blur gain $G_{motion}$ in Eq.~\ref{eq:motion_time_window}, iterative zero-intercept linear regression was applied. First, a digital twin of the experiment was created in the simulation, and the virtual camera was placed to generate synthetic images using Chrono::Sensor. The value of $G_{motion}$ was initialized to 1.0. After the synthetic images were generated, the synthetic transition angles were obtained using the same procedure as in the real experiments. Next, the ratio between the real and synthetic transition angles was computed for each condition, and then the ratios of all 18 conditions were averaged as the scaling factor used to update $G_{motion}$. This process was repeated twice until $G_{motion}$ converged to 1.283.

\begin{figure}[!t]
	\centering
	\includegraphics[width=0.6\columnwidth]{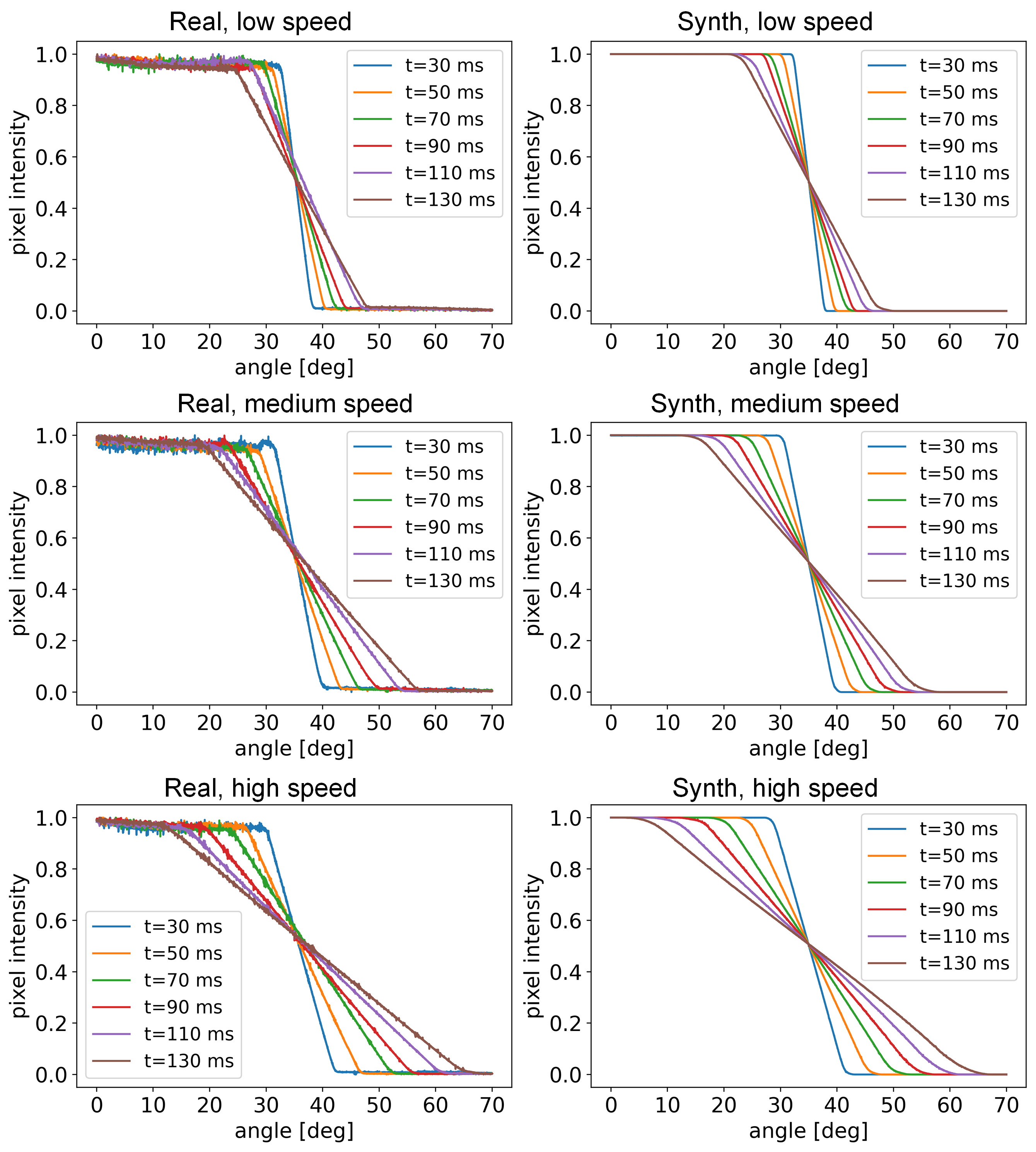}
	\caption{Comparison of transitional step responses between real photos (Real) and synthetic images (Synth) under different combinations of rotation speeds (high, medium, and low) and exposure times $t$ from 30 to 130 ms.}
	\label{fig:motion_blur_steps}
\end{figure}

\begin{figure*}[!t]
	\centering
	\includegraphics[width=1.0\columnwidth]{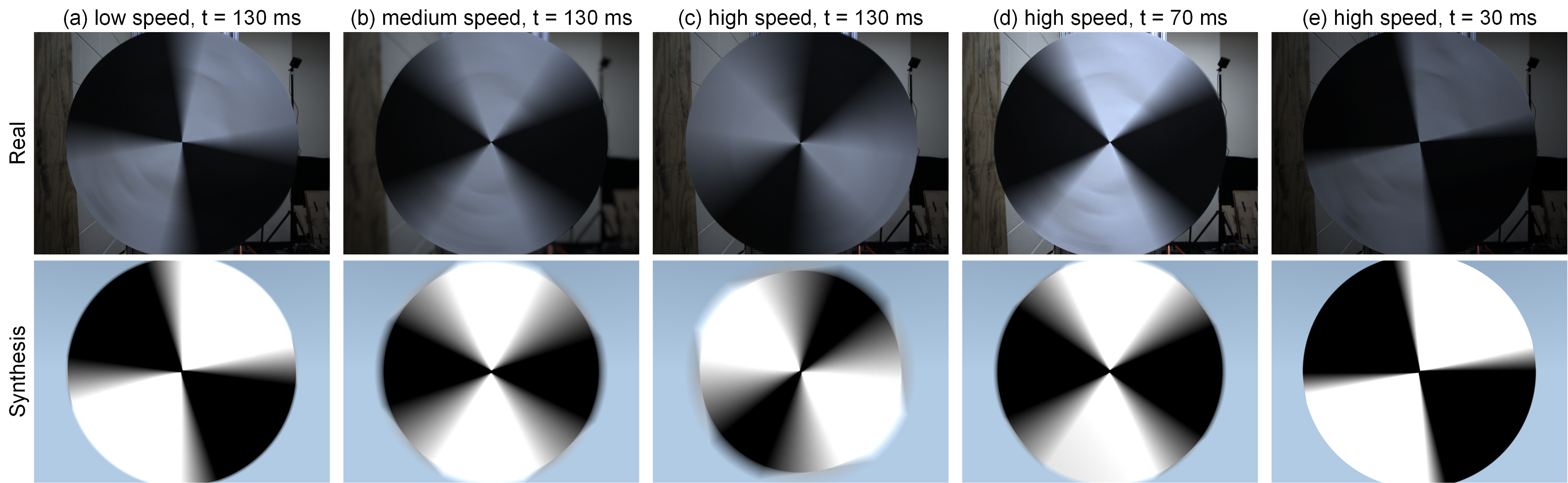}
	\caption{Pairs of real and synthetic pictures captured with different exposure times under different rotation speed. The brightness of some real photos was adjusted for visualization here.}
	\label{fig:motion_blur_examples}
\end{figure*}

The transitional step responses of pixel intensities over angles for all real and calibrated simulated conditions are qualitatively shown in Fig.~\ref{fig:motion_blur_steps}. The similar trends and consistent value scales between the synthetic and real curves indicate that the simulated motion blur closely replicates the real-world behavior. Moreover, quantitative comparisons of the transition angles between the real and calibrated synthetic results are presented in Table~\ref{table:motion_transitional_angles}. The results show that the magnitude and range of the synthetic transition angles closely match their real counterparts. The average error rate of the transition angles between the real and synthetic results across all the 18 conditions is 2.37\%. Several pairs of real photos and synthetic images after calibration are shown in Fig.~\ref{fig:motion_blur_examples}. The angular spans of the motion-blurred gradient sectors between black and white regions in the synthetic images closely resemble those in the corresponding real photos. This evidence further indicates that the simulated motion blur matches the real motion blur well.

\begin{table}[!t]
	\centering
	\caption{Transition angles (motion blur amount) of real photos and synthetic images.}
	\renewcommand{\arraystretch}{1.10}
	\begin{tabular}{|c|c|c|c|c|c|c|}
		\hline
		\multicolumn{7}{|c|}{Real}  \\
		\hline
		Speed \textbackslash{} $t$ & 30 & 50 & 70 & 90 & 110 & 130 \\
		\hline
		Low			& 4.63		& 7.70		& 10.49		& 13.68		& 16.24		& 19.87		\\
		\hline
		Medium		& 6.99		& 12.25		& 16.86		& 21.81		& 26.74		& 31.74		\\
		\hline
		High		& 10.14		& 16.81		& 23.75		& 30.23		& 37.78		& 42.89		\\
		\hline
		\multicolumn{7}{|c|}{Synthetic}  \\
		\hline
		Speed \textbackslash{} $t$ & 30 & 50 & 70 & 90 & 110 & 130 \\
		\hline
		Low			& 4.34		& 7.25		& 10.50		& 12.61		& 16.32		& 19.44		\\
		(err rate)	& (6.3\%)	& (5.8\%)	& (0.1\%)	& (7.8\%)	& (0.5\%)	& (2.2\%)	\\
		\hline
		Medium		& 7.24		& 12.05		& 16.78		& 21.58		& 26.31		& 31.22		\\
		(err rate)	& (3.6\%)	& (1.6\%)	& (0.5\%)	& (1.1\%)	& (1.6\%)	& (1.6\%)	\\
		\hline
		High		& 10.20		& 16.88		& 23.85		& 30.68		& 38.10		& 45.59		\\
		(err rate)	& (0.6\%)	& (0.4\%)	& (0.4\%)	& (1.5\%)	& (0.8\%)	& (6.3\%)	\\
		\hline
		\multicolumn{7}{l}{unit: [deg], $t$: exposure time, [ms], error rates were computed relative to} \\
		\multicolumn{7}{l}{the real transition angles.} \\
	\end{tabular}
	\label{table:motion_transitional_angles}
\end{table}

\subsubsection{Final calibrated synthetic image demonstration}
\label{subsubsec:cali_final}
\begin{figure*}[!t]
	\centering
	\includegraphics[width=1.0\columnwidth]{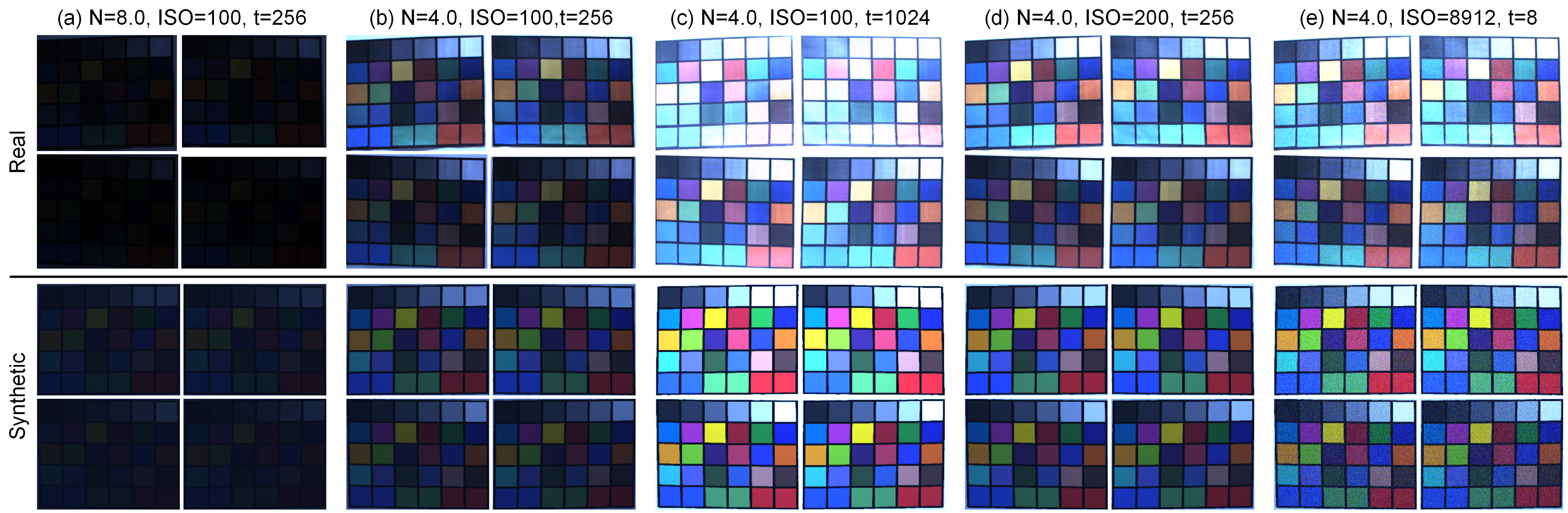}
	\caption{Comparison of the four MCCs cropped from the real photos (top) and synthetic images (bottom) under different camera settings (a) to (e), where $N$ denotes the aperture number and $t$ is the exposure time in milliseconds.}
	\label{fig:real_synth_cmpr}
\end{figure*}

With all layers in the proposed camera simulator calibrated, the final simulation results, incorporating all modeled optical artifacts, are demonstrated in Fig.~\ref{fig:real_synth_cmpr}. Figure~\ref{fig:real_synth_cmpr} compares the four cropped MCCs from the respective real and synthetic pictures. The comparison between real photos and synthetic images illustrates that the calibrated camera simulator successfully replicates key optical effects, including vignetting, exposure variations, and noise patterns.

Exposure variations align well between real and synthetic pictures, demonstrating the model's accuracy in simulating changes in brightness. Noise behavior is another crucial aspect. As shown in Fig.~\ref{fig:real_synth_cmpr}(e), high ISO settings exhibit noticeable noise levels in both real and synthetic pictures. The Poisson-Gaussian noise model successfully replicates the real-world noise characteristics.


\section{Real-to-Sim Virtual Environment Creation Study}
\label{sec:exp_steup}
Drawing on the vision illustrated in Fig.~\ref{fig:concat_NeRF}, we integrate the proposed differentiable physics-based camera simulator with a differentiable renderer to optimize mesh geometry and material textures for virtual environment creation. This application is motivated by the robotics challenge introduced at the beginning of this paper: \textit{How can we build a high-fidelity digital twin of a real off-road environment to evaluate an AGV navigation policy in simulation?} A straightforward approach involves capturing real-world photographs of the scene, generating a digital twin (i.e., a virtual environment) via inverse rendering, integrating it with a vehicle dynamics model in a multi-physics simulator (we use Project Chrono \cite{chronoOverview2016} in this work), and running simulations to validate the performance of the navigation policy \cite{shen2024driveenv-nerf,murali2024learning}.

The differentiable renderer used to implement this procedure was NVDiffRecMC \cite{hasselgren2022shape}. Instead of the other state-of-the-art method, Gaussian Splatting \cite{kerbl2023gaussian} combined with SuGaR \cite{guedon2024sugar}, NVDiffRecMC has several advantages that meet our requirements. Unlike Gaussian Splatting with SuGaR, which outputs an albedo texture and a geometric mesh without normals, NVDiffRecMC generates multiple material textures essential for off-the-shelf shaders and Chrono::Sensor, including albedo, normals, roughness, and metallic textures. In addition, NVDiffRecMC employs the simplified Disney BRDF \cite{burley2012physically} in its shader, which is also used in Chrono::Sensor and can ensure consistency between inverse rendering and multi-physics simulation.

Furthermore, NVDiffRecMC can cast shadows based on ray-tracing under different illumination maps, while Gaussian Splatting is rasterization-based and lacks shadow modeling. This capability is particularly important in outdoor environments, such as lunar terrains, where obstacles cast dynamic and sometimes elongated shadows due to variations in sunlight. However, such shadows can severely obscure surface details and degrade texture optimization in shadowed regions. To mitigate this loss of surface detail, photos should be captured at different times of the day to ensure that regions occluded by shadows in one illumination condition are visible under another. Unlike the original NVDiffRecMC, which optimizes a single illumination map, we used multiple fixed illumination maps. These maps were pre-calibrated and aligned with the lighting probes in simulation by comparing the reflected colors of a reference cube with known surface materials rendered in both NVDiffRecMC and Chrono::Sensor, ensuring that the lighting was consistent between the two renderers so that textures optimized in NVDiffRecMC appear correct when rendered in Chrono::Sensor.

We evaluate this real-to-sim pipeline on two complementary test beds, described in the following subsections together with the evaluation metrics, ablation configurations, and results. The first test bed consists of three simulated scenes, where the ground-truth geometry and materials are known exactly, enabling a controlled study of how each camera-model component affects reconstruction; the three scenes probe general appearance (Scene~1), geometry recovery (Scene~2), and material recovery (Scene~3). The second consists of two real-world scenes captured with the calibrated camera, which validate the approach on real photographs and provide the environment for the AGV demonstration in Sec.~\ref{sec:AGV_demo}.

\subsection{Simulated scene setup}
\label{subsec:sim_scene_setup}
To evaluate the proposed camera simulator for novel image synthesis, three simulated scenes were created to generate synthetic photographs. Each of our scenes contained more than ten objects distributed throughout the environment. This arrangement better highlighted the effects of defocus blur, particularly when the camera was focused on objects at varying distances (near or far). Moreover, the spatial distribution of objects reflects real-world outdoor environments, where obstacles are scattered and the AGV must plan its trajectory accordingly.

\begin{figure}[!t]
	\centering
	\includegraphics[width=0.6\columnwidth]{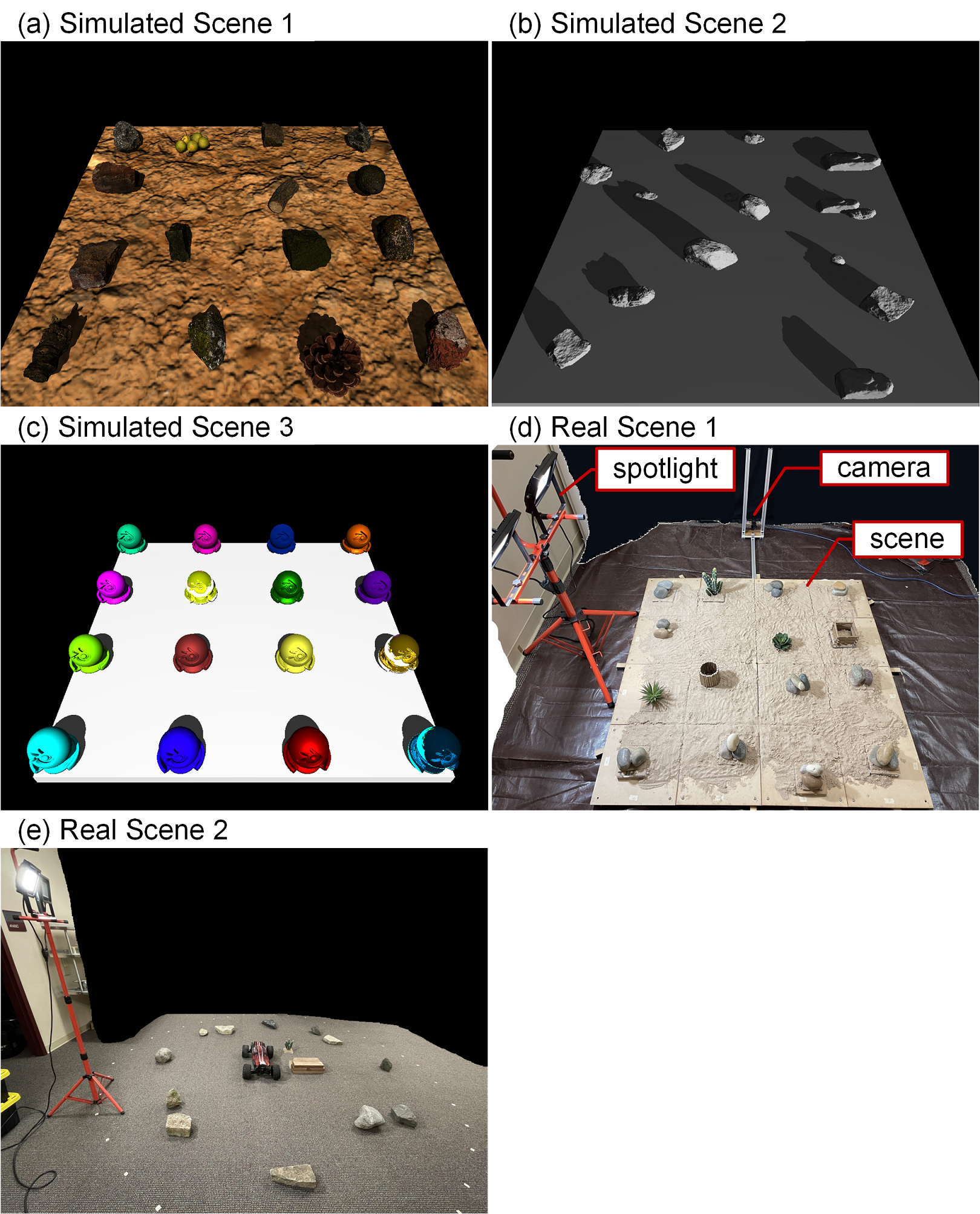}
	\caption{Illustrations of the simulated scenes -- (a) to (c), and of the real-world scenes -- (d) and (e). For each scene, the downward direction is the global +X axis, rightward direction is the +Y axis, and upward direction of the ground is the +Z axis.}
	\label{fig:scene_demo}
\end{figure}

Each scene was divided into two components: objects and ground. During virtual environment construction, illumination was provided by a directional light simulating sunlight and a weak ambient light. Several objects and a ground plane were placed in the scene, and both a pinhole camera and a depth camera from Chrono::Sensor \cite{asherSensorSimulation2021} were used to generate the scene radiance and object depth maps, respectively, at various camera poses.
The depth maps were used to compute object region-of-interest (ROI) masks and defocus blur weight matrices, as defined in Eq.~\ref{eq:defocus_blur}(b). Non-object pixels in the radiance images were masked out, and the calibrated, PyTorch-based camera simulator was then applied with varying camera settings and corresponding blur weight matrices to synthesize the final images.
Scenes 1 and 3 were arranged on a square ground with a side length of 1.32~m, while Scene 2 used a 1.5~m square ground. Figure~\ref{fig:scene_demo} illustrates the virtual environments.

\begin{itemize}[leftmargin=*]
	\item Scene 1, as shown in Fig.~\ref{fig:scene_demo}(a), simulated an off-road terrain, containing 15 natural objects, including rocks, wooden logs, and plant fruits. These objects were uniformly distributed. The sun elevation was set to 45 degrees. This scene featured a variety of object shapes and plentiful colors.
	
	\item Scene 2, shown in Fig.~\ref{fig:scene_demo}(b), was designed to mimic a lunar environment by replicating the rock distribution, rock sizes, and sun elevation angle (8$^\circ$) from Terrain~1 in the POLAR dataset \cite{wong2017polar}. The scene's radiance was generated using the Hapke BRDF model \cite{Hapke2012,sato2014resolved}, which simulates lunar light scattering based on empirical measurements of the Moon's surface. The low solar elevation -- characteristic of the lunar South Pole regions of interest in the ongoing NASA Artemis program -- posed reconstruction challenges, as the shaded sides of rocks lost substantial surface detail. The scene included objects of various shapes but composed of a single material, focusing the evaluation on mesh optimization.
	
	\item As shown in Fig.~\ref{fig:scene_demo}(c), Scene~3 consisted of 16 standard Blender pillars \cite{materials_pillar} arranged in a uniform grid, a configuration commonly used in NeRF evaluations. Each pillar was randomly assigned a base color, and its roughness and metallic values were selected at random from all cross-combinations of the set $\{0,\; 0.33,\; 0.67,\; 1\}$. The sun elevation was set to 45 degrees. This scene featured objects with varying surface materials but identical shape, focusing the evaluation on material optimization.
\end{itemize}

For each simulated scene, a camera was positioned on the upper hemisphere around the scene, capturing synthetic photos from various polar and azimuth angles and always pointing toward the scene center. The polar angles (where the upward direction of the ground is the +Z axis at a polar angle of 0 degrees) were set to 40, 55, 70, and 85 degrees. For high elevations (40 and 55 degrees), where the depth of field was shallow, the camera was focused at the center of the scene. For lower elevations (70 and 85 degrees), where the depth of field was deeper, the camera alternated its focus between near and far distances at each pose, presenting obvious defocus blur artifacts. Small aperture numbers of 2.0 and 1.6 were used for high and low elevation angles, respectively, to enhance the defocus blur effect. Three exposure times, 128, 256, and 512 milliseconds, were set to capture photos with different brightness levels. The ISO was fixed at 100, and the focal length was fixed at 5 mm, as it was a prime lens.

To introduce variation in illumination, the sun azimuth angles were set to $0^\circ$, $90^\circ$, and $180^\circ$ for Scenes~1 and~3, and to $45^\circ$, $135^\circ$, and $315^\circ$ for Scene~2. This resulted in 1,377 synthetic photos per scene. The training and testing sets were split by camera poses in a 2:1 ratio.

\subsection{Real scene setup}
\label{subsec:real_scene_setup}
To validate the method on real photographs, two physical experimental scenes were constructed and captured with the real camera, as shown in Fig.~\ref{fig:scene_demo} (d) and (e). These real scenes were intended to imitate real-world off-road terrain. In Real Scene 1, the ground was a 1.32 m $\times$ 1.32 m high-density fiberboard, with 15 objects uniformly distributed on it. These objects varied in natural material compositions, including piled river rocks, wooden constructions, and fake succulent plants (agave, cactuses, and echeveria). Each object had an approximate width of 12 cm. Sand was scattered across the board to mimic outdoor terrain conditions. To control photo capture, a 4-foot-long aluminum framing rail was installed with a camera-supporting frame mounted on it. The rail allowed the entire assembly to rotate around the center of the ground plane beneath the fiberboard. The camera-supporting frame could be translated along the rail and adjusted in height and pitch angle, enabling precise positioning at desired viewing poses. The scene was illuminated by a dimmable LED spotlight (10{,}000~lm, 5000~K) operated at 10\% power to simulate sunlight, with the room's ceiling light serving as ambient illumination. The illumination maps in this setup were calibrated by comparing the reflected colors of a stack of white printing paper, whose surface material properties were known, between NVDiffRecMC renderings and real photographs.

Real Scene 2 was configured similarly to Real Scene 1. To emulate a larger real-world off-road terrain, 12 rocks were distributed within a circular region with a diameter of 3.0~m. Each rock had an average size of 30~cm. A scaled-down off-road vehicle, a pile of wooden strips, and a miniature cactus were placed at the center of the circle. The scene was illuminated by the same LED spotlight used in Real Scene 1, operated at 20\% power to simulate sunlight, and the room's ceiling light served as ambient illumination.

The camera configurations were similar to those used in the simulated scenes. For Real Scene 1, the camera was placed at four polar angles: 40, 55, 70, and 85 degrees. For each polar angle, 24, 18, 18, and 18 uniformly spaced azimuth angles were used, respectively. For Real Scene 2, the first polar angle was changed from 40 degrees to 45 degrees, and 18 uniformly spaced azimuth angles were used at this elevation. At the first two higher camera elevations, the camera was focused at the scene center with an aperture number of 2.8. At the last two lower elevation angles, the camera was alternately focused at the near 1/8 and far 7/8 scene ranges, using a lower aperture number of 2.0 to present defocus blur effects more explicitly. Photographs were taken using two exposure times (128 and 256~ms) for Real Scene 1 and three exposure times (80, 120, and 160~ms) for Real Scene 2, respectively. The ISO was fixed at 100, and the focal length was 5~mm.

For data post-processing, VGGT~\cite{wang2025vggt} and COLMAP~\cite{schonberger2016structure,schoenberger2016mvs} were used to estimate precise camera poses for all photos. The ROI and depth maps of the scene were generated through a multi-step process. First, SuGaR~\cite{guedon2024sugar} was used to reconstruct a 3D mesh of the scene from the captured photos. The mesh was then segmented into a collection of object meshes and a ground surface mesh (the latter will be used in Sec.~\ref{sec:AGV_demo}). Next, the object meshes were imported into Mitsuba~\cite{Mitsuba} to render the ROI and depth maps. The ROI map was applied to mask non-object pixels in the real photos, while the depth map was used to compute the defocus blur weight matrices required for Eq.~\ref{eq:defocus_blur}. The training and testing sets were split by camera poses. For Scene 1, 234 real photos were used for training and 112 photos were reserved as ground-truth for testing. For Scene 2, 375 real photos were used for training and 186 photos were used as ground-truth for testing.

\subsection{Metrics, baseline, and ablation study setting}
\label{subsec:metrics}
To evaluate the proposed image synthesis and 3D reconstruction approaches, three widely used validation metrics were employed: Peak Signal-to-Noise Ratio (PSNR), Structural Similarity Index Measure (SSIM) \cite{wang2004image}, and Learned Perceptual Image Patch Similarity (LPIPS) \cite{zhang2018unreasonable}.

Since this work primarily focuses on objects in masked images, where the objects are spatially separated and occupy only a small portion of each image, the PSNR calculation was modified to ensure fairness. Specifically, instead of computing the mean squared error over all pixels, it was computed only over the effective pixels (i.e., those corresponding to object regions). That is,
{\small
	\begin{equation}
		\label{eq:modified_PSNR}
		PSNR = -10 \cdot \log_{10} \left( \sum \limits_{i=1}^{h} \sum \limits_{j=1}^{w} \frac{\left( y_{ij} - \hat{y}_{ij} \right) ^2}{\text{number of effective pixels}} \right) \; ,
\end{equation}}
where $y_{ij}$ and $\hat{y}_{ij}$ denote the predicted and ground-truth pixel values, respectively, both normalized to the range $[0, 1]$, and $h$ and $w$ represent the image height and width in pixels.

To quantify the impact of various components of the proposed camera simulator on its performance, baseline and ablation studies were conducted by comparing different configurations of the simulator. The following four cases were examined. In all cases, the noise-adding layer was disabled, except for the dark current, and lens distortion was ignored:
\begin{itemize}[leftmargin=*]
	\item \textbf{full camera}. This setup activated all components of the camera simulator after NVDiffRecMC, incorporating all optical effects.
	
	\item \textbf{w/o defocus blur}. This setup deactivated the defocus blur component to evaluate its impact on optimization and synthetic image quality.
	
	\item \textbf{w/o exposure-related}. This setup disabled all color- and brightness-related layers, including vignetting, aggregation, and the CRF, leaving only the defocus blur component activated. The goal was to assess the impact of color and brightness modulation.
	
	\item \textbf{NVDiffRecMC only \cite{hasselgren2022shape}}. This baseline setup removed the entire camera simulator from the pipeline, relying solely on NVDiffRecMC for optimization.
\end{itemize}

For the last two setups, to account for brightness differences caused by varying exposure times, the digital values of the images were adjusted by multiplying them by the ratio of the exposure time to 256~ms, with 256~ms taken as the baseline.

Regarding the choice of the defocus blur kernel in the \textit{full camera} and \textit{w/o exposure-related} setups, only the Gaussian kernel was adopted for the simulated scenes, since the simulated ground-truth photos were generated using the proposed camera simulator, and the Gaussian kernel performed better after calibration in Sec.~\ref{subsubsec:defocus_cali}. In contrast, for the real-world scenes, both the Gaussian and Uniform kernels were used for comparison.

Each scene was optimized in two passes. For each condition setup, each simulated scene was optimized for 5{,}000 iterations per pass, while each real scene was optimized for 20{,}000 iterations per pass. Additional hyperparameter settings are provided in Appendix~\ref{sec:appndx_opt}, with all remaining parameters kept at their default values in NVDiffRecMC. All optimization procedures were performed on a machine equipped with an Intel i7-13700K CPU, 32~GB of RAM, and an NVIDIA GeForce RTX~4080 GPU with 16~GB of VRAM.

\subsection{Baseline and ablation study results}
\label{subsec:results}
For both simulated and real scenes, training-set photos were used to recover mesh geometry and material textures for 3D scene reconstruction. This is where the differentiable property of the camera model was exploited. After reconstruction, synthetic images were rendered using the optimized mesh and textures, and then compared to corresponding ground-truth photos in the testing set using PSNR, SSIM, and LPIPS to assess reconstruction quality.

The optimization procedure took approximately 50 minutes for each simulated scene setup and about three hours for each real scene setup, respectively. These computational expenses are substantially lower than those reported in NeRFocus \cite{wang2022nerfocus}. NeRFocus also aimed to model physics-based defocus blur with NeRF and used a single NVIDIA V100 GPU, but required over 50 hours to complete a single training session.

Quantitative evaluation results of the baseline and ablation studies for both the simulated and real-world scenes are reported in Table~\ref{tab:PSNR_cmpr} (PSNR), Table~\ref{tab:SSIM_cmpr} (SSIM), and Table~\ref{tab:LPIPS_cmpr} (LPIPS). Across all simulated scenes, using the full camera simulator consistently yielded the best performance. This setup achieved the highest PSNR and SSIM scores and the lowest LPIPS values. The performance difference between using the full simulator and disabling the defocus blur layer was relatively small, suggesting that defocus blur has limited impact on the quantitative metrics considered. A similar trend was observed when comparing the setups that removed the exposure-related layers versus those that omitted the entire camera simulator. Notably, the significant drop in performance when deactivating the exposure-related layers underscores the substantial influence of color- and brightness-related camera effects on quantitative evaluation.

For the real-world scenes, the configurations without the defocus blur layer achieved the highest PSNR values in Table~\ref{tab:PSNR_cmpr}. Since PSNR is the most important index, which directly measures pixel-wise reconstruction errors, this result suggests that omitting defocus blur yields the closest pixel-level agreement with the ground-truth photos. In contrast, the SSIM and LPIPS results were similar among the full camera simulator using Gaussian-kernel defocus blur, without the defocus blur layer, and without the camera simulator. Performance dropped noticeably when only the exposure-related layers were deactivated, further indicating that exposure modulation plays a more critical role in quantitative performance. In addition, the overall performance on the real experimental scenes was lower than that on the simulated scenes, highlighting the challenges of domain adaptation and sim-to-real transfer. This discrepancy suggests that the physical camera modeling and calibration process could be further improved. Moreover, among the setups with the defocus blur layer activated, all evaluation metrics obtained using the Gaussian blur kernel were better than those obtained using the Uniform blur kernel. This result further indicates that the Gaussian kernel better fits the defocus blur artifact.

\begin{figure*}[!t]
	\centering
	\includegraphics[width=1.0\columnwidth]{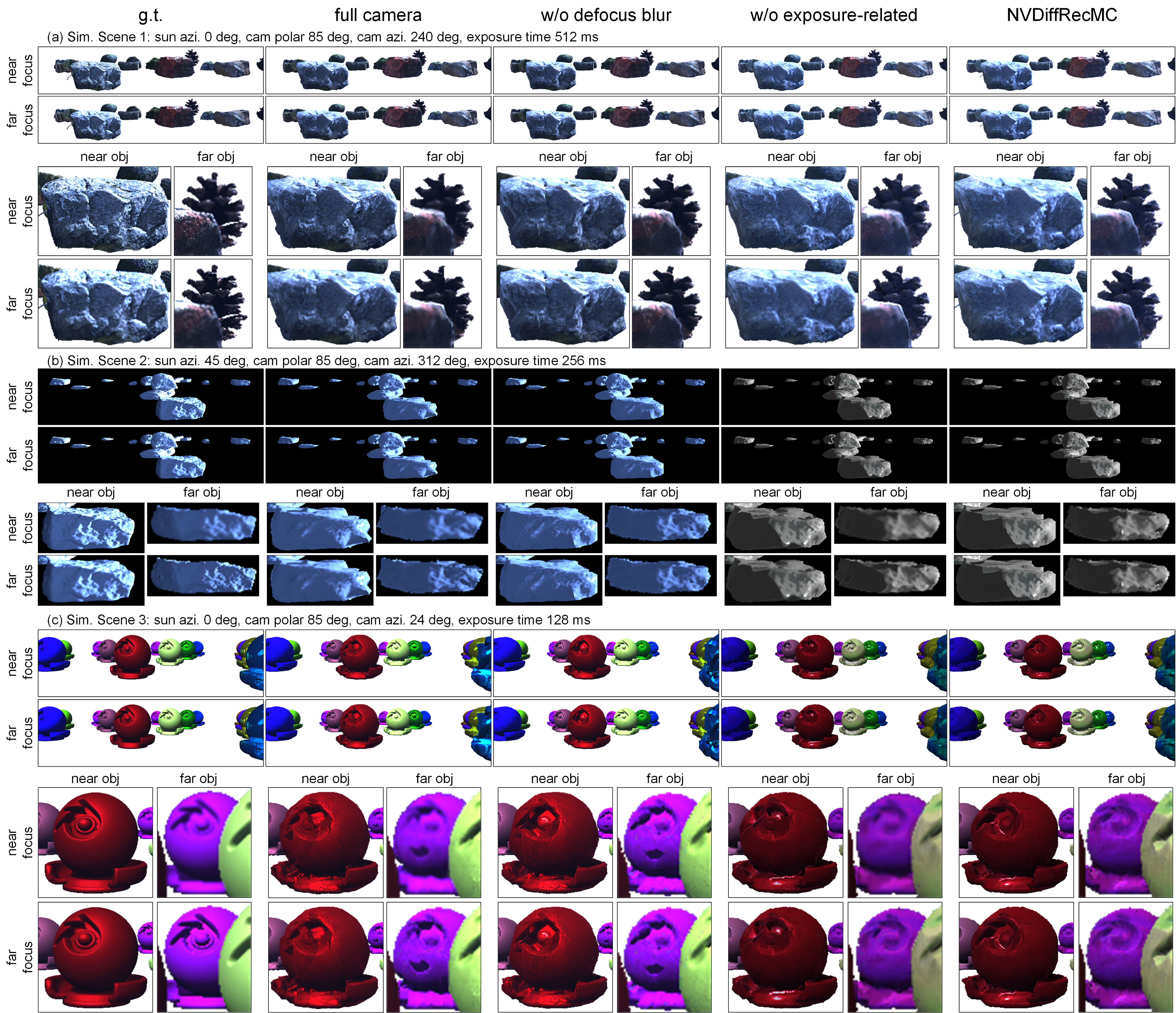}
	\caption{Comparison of defocus blur effects from testing viewpoints in Simulated Scenes 1 to 3 (from (a) to (c)) among different ablation study conditions (from left to right: ground-truth, with full camera simulator, without the defocus blur layer, without exposure-related layers, and the baseline only using NVDiffRecMC\cite{hasselgren2022shape}), and their zoomed-in clips of the near and far objects for clearer demonstration. For each scene, the upper row is near-focused and the lower row is far-focused.}
	\label{fig:sim_defocus_cmpr}
\end{figure*}

\begin{figure*}[!t]
	\centering
	\includegraphics[width=1.0\columnwidth]{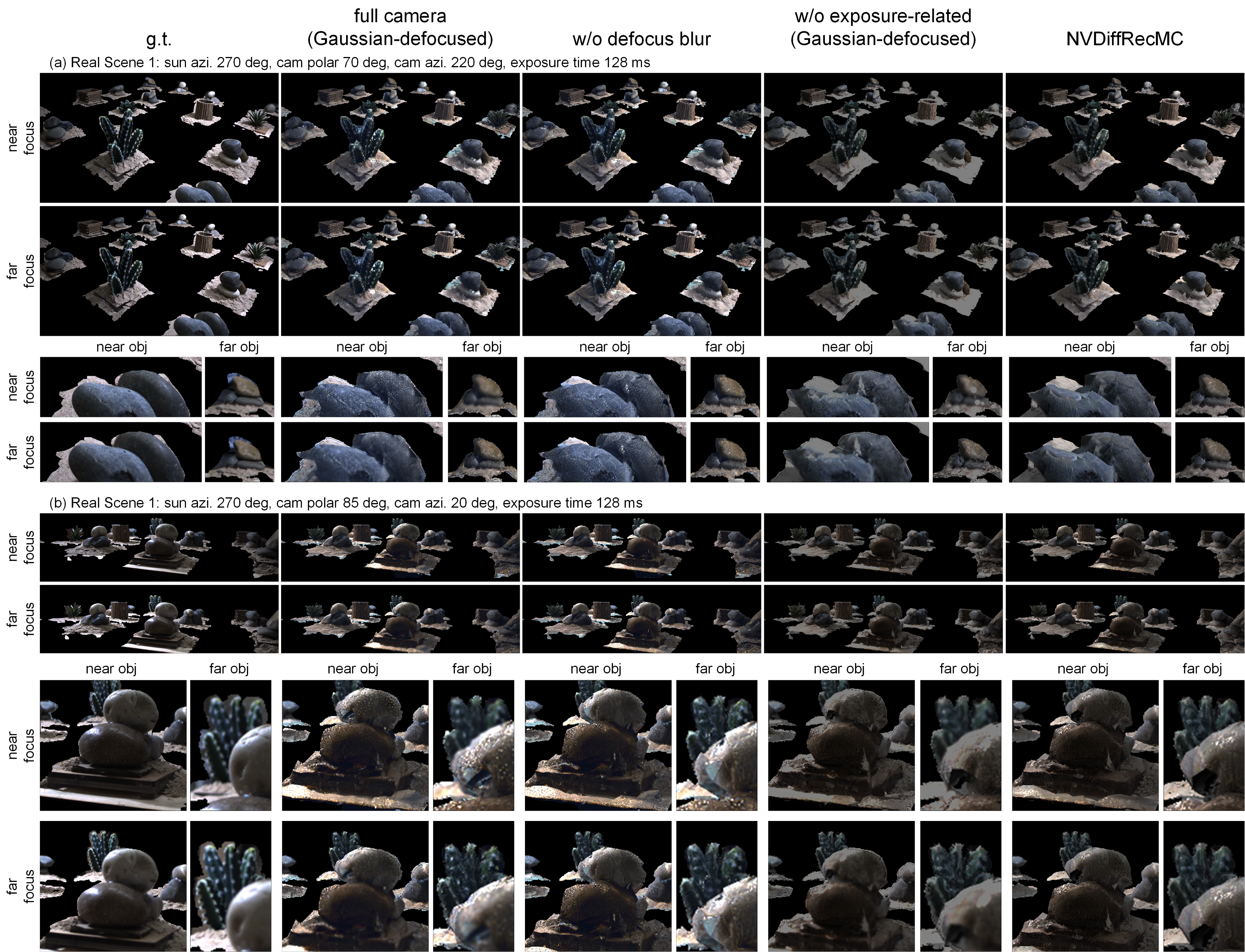}
	\caption{Comparison of defocus blur effects from two testing viewpoints ((a) and (b)) of Real Scene 1 among different baseline and ablation study conditions (same condition order as Fig.~\ref{fig:sim_defocus_cmpr}), with zoomed-in clips of the near and far objects for clearer demonstration. For each viewpoint, the upper row is near-focused and the lower row is far-focused.}
	\label{fig:real01_defocus_cmpr}
\end{figure*}

\begin{figure*}[!t]
	\centering
	\includegraphics[width=1.0\columnwidth]{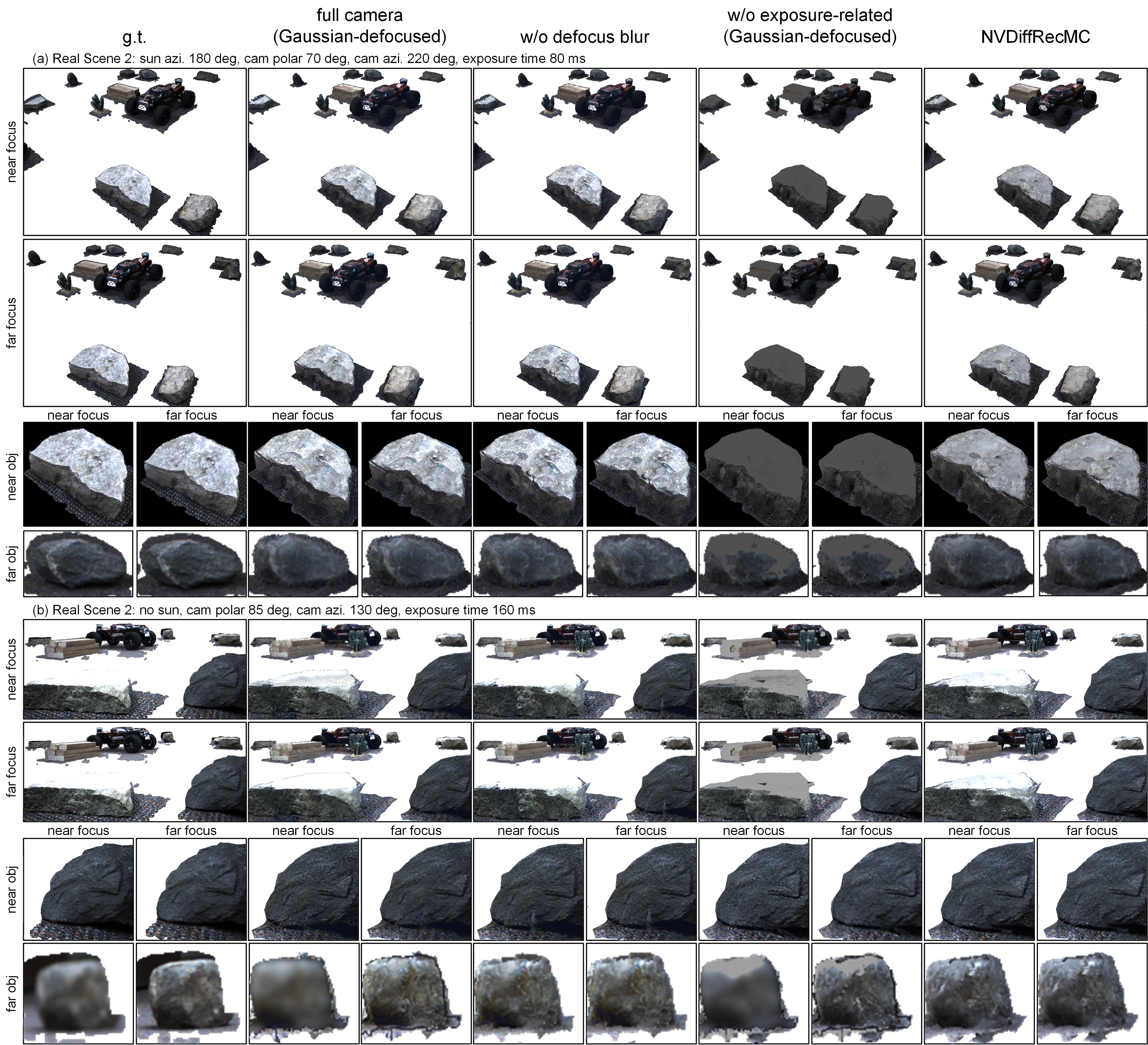}
	\caption{Comparison of defocus blur effects from two testing viewpoints ((a) and (b)) of Real Scene 2 among different baseline and ablation study conditions (same condition order as Fig.~\ref{fig:sim_defocus_cmpr}), with zoomed-in clips of the near and far objects for clearer demonstration. Notice that the ground-truth photos and configurations in (b) are not part of the testing set; they are additional data collected only for defocus blur visualization. For image synthesis in (b), the function used to calculate the defocus blur diameter $D_{ij}$ in Eq.~\ref{eq:defocus_blur_diameter} is modified by: adding a dead-zone radius, $b = 0.25\;m$, such that pixels of depths within the dead-zone range centered at the focus distance have zero defocus blur diameter.}
	\label{fig:real02_defocus_cmpr}
\end{figure*}

Figures~\ref{fig:sim_defocus_cmpr}, \ref{fig:real01_defocus_cmpr}, and \ref{fig:real02_defocus_cmpr} present qualitative comparisons of defocus blur effects from testing viewpoints of the simulated and real scenes. In the rows of zoomed-in images, the defocus blur effect is slight on near objects when the camera is far-focused. Conversely, when focusing on near objects, the defocus blur effects of far objects appear more evident. However, the pixel contribution from far objects is small due to their reduced size in the image. This likely explains why the defocus blur effect does not significantly degrade evaluation metrics but can only be distinguished qualitatively.

Qualitative comparisons also show that defocus blur artifacts are clearly visible in all scenes when the defocus blur component is activated, indicating that the defocus blur component successfully reproduces realistic camera behavior. In contrast, when the component is disabled, regardless of focus distance, the appearances of both near and far objects converge to sharply rendered surfaces with very little blur, demonstrating a lack of depth-dependent defocus in the absence of the blur model.

\begin{figure*}[!t]
	\centering
	\includegraphics[width=0.85\columnwidth]{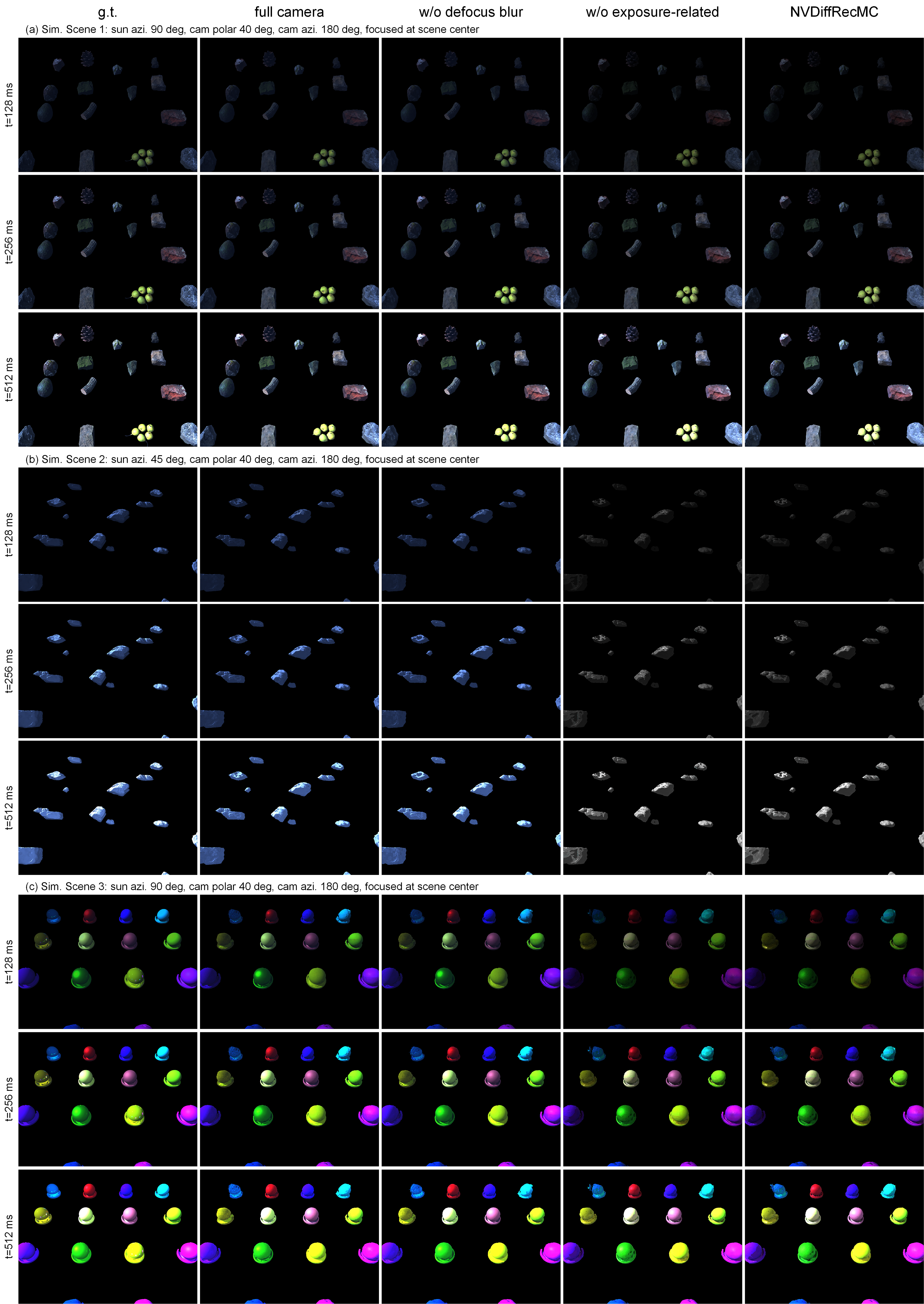}
	\caption{Comparison of different exposure times from testing viewpoints of Simulated Scenes 1 to 3 (from (a) to (c)) among different baseline and ablation study conditions (same condition order as Fig.~\ref{fig:sim_defocus_cmpr}). For each scene, the exposure times from top to bottom are 128, 256, and 512 milliseconds, respectively.}
	\label{fig:sim_expsr_cmpr}
\end{figure*}

\begin{figure*}[!t]
	\centering
	\includegraphics[width=1.0\columnwidth]{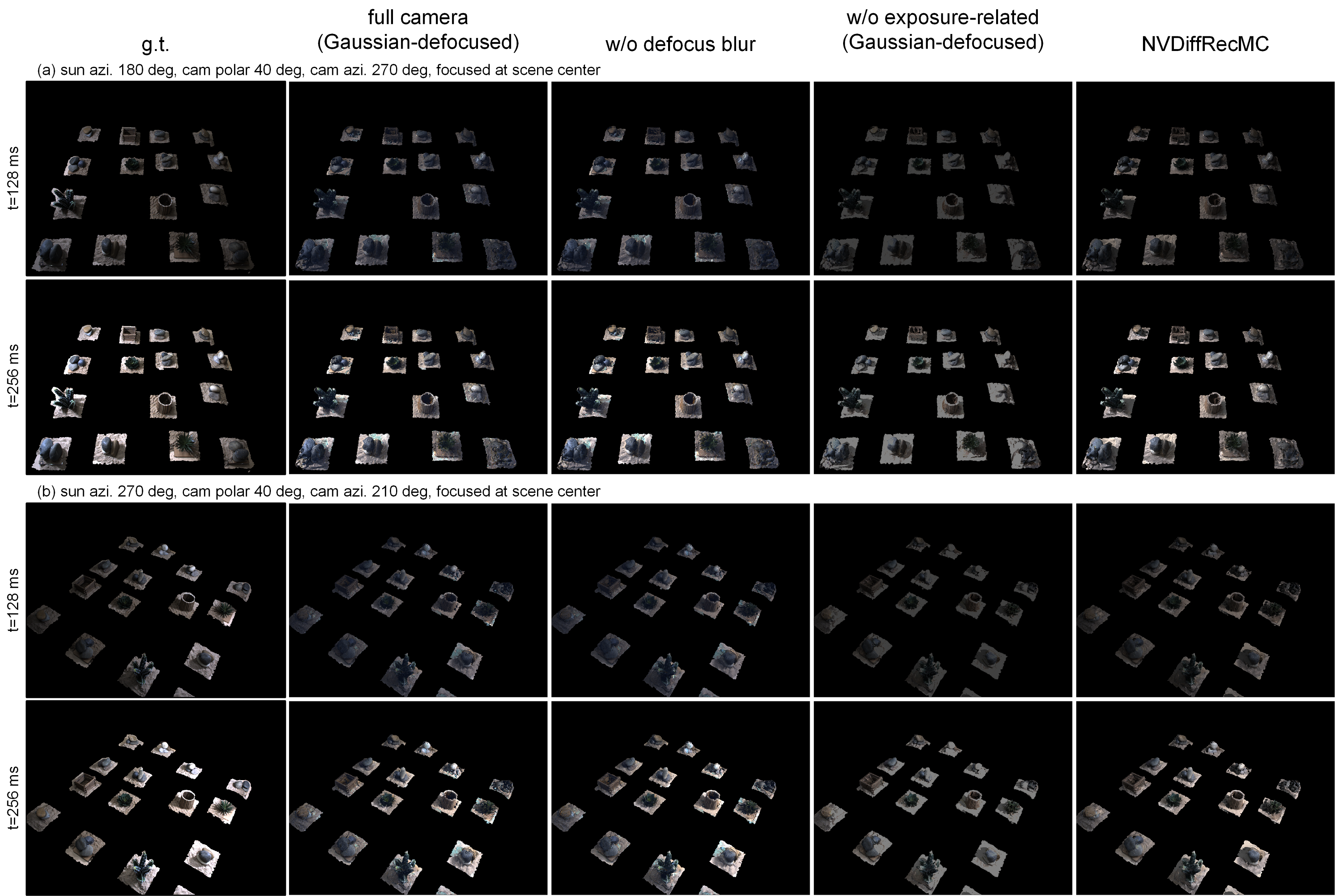}
	\caption{Comparison of different exposure times from two testing viewpoints ((a) and (b)) of Real Scene 1 among different baseline and ablation study conditions (same condition order as Fig.~\ref{fig:sim_defocus_cmpr}). For each viewpoint, the exposure times from top to bottom are 128 and 256 milliseconds, respectively.}
	\label{fig:real01_expsr_cmpr}
\end{figure*}

\begin{figure*}[!t]
	\centering
	\includegraphics[width=1.0\columnwidth]{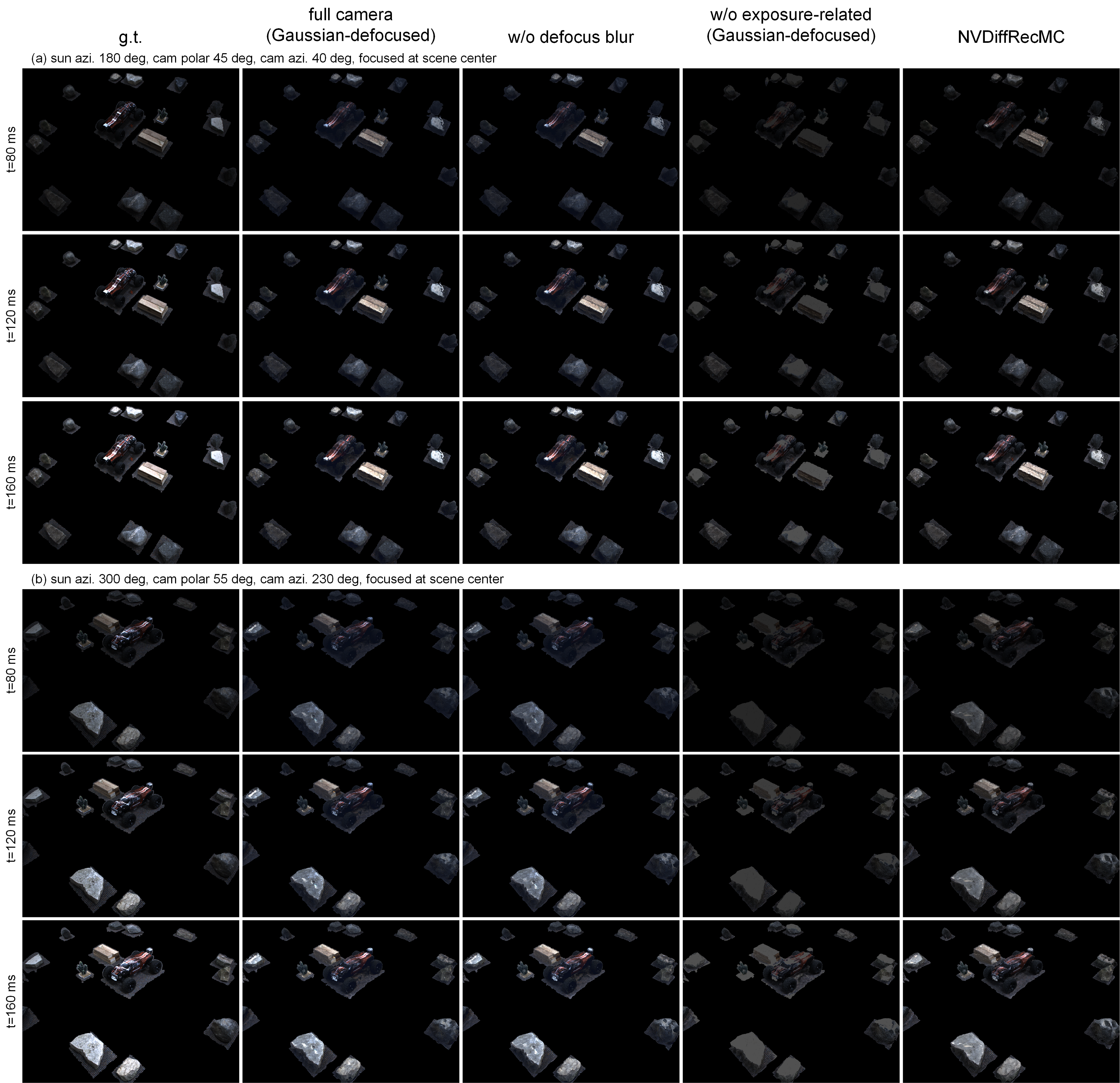}
	\caption{Comparison of different exposure times from two testing viewpoints ((a) and (b)) of Real Scene 2 among different baseline and ablation study conditions (same condition order as Fig.~\ref{fig:sim_defocus_cmpr}). For each viewpoint, the exposure times from top to bottom are 80, 120, and 160 milliseconds, respectively.}
	\label{fig:real02_expsr_cmpr}
\end{figure*}

Figures~\ref{fig:sim_expsr_cmpr}, \ref{fig:real01_expsr_cmpr}, and \ref{fig:real02_expsr_cmpr} present qualitative comparisons of exposure variations from testing viewpoints for the simulated and real-world scenes. The results show that all ablation setups were able to reproduce exposure variations to some extent, even when the exposure-related components were deactivated and the images were adjusted solely by scaling according to exposure time ratios. This suggests that the dark current in the noise component and the bias term in the CRF component have only a minor impact on the visual output.

However, subtle differences still existed when the exposure-related components were deactivated. In Simulated Scene 2, the color tones were significantly biased when the exposure-related layers were removed, as the ground-truth images appeared more blue due to the calibrated channel sensitivity in the camera simulator. This explains the large performance drop in PSNR for Scene 2 in Table~\ref{tab:PSNR_cmpr}. In addition, in Simulated Scene 3, a close examination of the green pillar in the third row from the top and second column from the left reveals that the specular reflectance was not rendered as well as the others when the exposure-related components were deactivated. These observations confirm that the exposure-related components in the camera simulator are important for texture optimization and color reproduction.

To conclude, the defocus blur component primarily enhances perceptual realism in qualitative observations, while the exposure-related components play a critical role in improving quantitative evaluation metrics. Therefore, using the full camera simulator during optimization is essential: the defocus blur component is necessary to replicate realistic depth-of-field effects, and the exposure-related components are required to achieve strong performance on quantitative metrics.

\begin{table}[!t]
	\small
	\centering
	\caption{Comparison of PSNR among different conditions.}
	\renewcommand{\arraystretch}{1.20}
	\begin{tabular}{|c|c|c|c|c|c|}
		\hline
		PSNR [dB] $\uparrow$	& Sim. 1			& Sim. 2			& Sim. 3			& Real 1			& Real 2 \\
		\hline
		\begin{tabular}[c]{@{}c@{}}
			full camera\\(Gaussian)
		\end{tabular}			& \textbf{11.326}	& \textbf{8.547}	& \textbf{9.846}	& 8.364				& 9.493 \\
		\hline
		\begin{tabular}[c]{@{}c@{}}
			full camera\\(Uniform)
		\end{tabular}			& --				& --				& --				& 8.193				& 9.336 \\
		\hline
		\begin{tabular}[c]{@{}c@{}}
			w.o. defocus
		\end{tabular}			& 11.304			& 8.543				& 9.789				& \textbf{8.528}	& \textbf{9.552} \\
		\hline
		\begin{tabular}[c]{@{}c@{}}
			w.o. expsr\\(Gaussian)
		\end{tabular} 			& 11.090			& 4.988				& 8.776				& 7.920				& 8.407	\\
		\hline
		\begin{tabular}[c]{@{}c@{}}
			w.o. expsr\\(Uniform)
		\end{tabular} 			& --				& --				& --				& 7.790				& 8.338	\\
		\hline
		\begin{tabular}[c]{@{}c@{}}
			NVDiffRecMC
		\end{tabular}  			& 11.074			& 5.019				& 8.697				& 8.244				& 9.348	\\
		\hline
		\multicolumn{6}{l}{The bold number marks the best value among all conditions.} \\
	\end{tabular}
	\label{tab:PSNR_cmpr}
\end{table}

\begin{table}[!t]
	\small
	\centering
	\caption{Comparison of SSIM among different conditions.}
	\renewcommand{\arraystretch}{1.20}
	\begin{tabular}{|c|c|c|c|c|c|}
		\hline
		SSIM $\uparrow$			& Sim. 1			& Sim. 2			& Sim. 3			& Real 1			& Real 2 \\
		\hline
		\begin{tabular}[c]{@{}c@{}}
			full camera\\(Gaussian)
		\end{tabular}			& \textbf{0.950}	& \textbf{0.967}	& \textbf{0.944}	& \textbf{0.876}	& 0.858	\\
		\hline
		\begin{tabular}[c]{@{}c@{}}
			full camera\\(Uniform)
		\end{tabular}			& --				& --				& --				& 0.874				& 0.854 \\
		\hline
		\begin{tabular}[c]{@{}c@{}}
			w.o. defocus
		\end{tabular}			& \textbf{0.950}	& 0.966				& 0.941				& \textbf{0.876}	& \textbf{0.859} \\
		\hline
		\begin{tabular}[c]{@{}c@{}}
			w.o. expsr\\(Gaussian)
		\end{tabular} 			& 0.947				& 0.954				& 0.931				& 0.875				& 0.852 \\
		\hline
		\begin{tabular}[c]{@{}c@{}}
			w.o. expsr\\(Uniform)
		\end{tabular} 			& --				& --				& --				& 0.872				& 0.851 \\
		\hline
		\begin{tabular}[c]{@{}c@{}}
			NVDiffRecMC
		\end{tabular}  			& 0.946				& 0.954				& 0.929				& \textbf{0.876}	& \textbf{0.859} \\
		\hline
		\multicolumn{6}{l}{The bold number marks the best value among all conditions.} \\
	\end{tabular}
	\label{tab:SSIM_cmpr}
\end{table}

\begin{table}[!t]
	\small
	\centering
	\caption{Comparison of LPIPS among different conditions.}
	\renewcommand{\arraystretch}{1.20}
	\begin{tabular}{|c|c|c|c|c|c|}
		\hline
		LPIPS $\downarrow$		& Sim. 1			& Sim. 2			& Sim. 3			& Real 1			& Real 2 \\
		\hline
		\begin{tabular}[c]{@{}c@{}}
			full camera\\(Gaussian)
		\end{tabular}			& \textbf{0.0507}	& \textbf{0.0364}	& \textbf{0.0734}	& 0.1091			& 0.1194 \\
		\hline
		\begin{tabular}[c]{@{}c@{}}
			full camera\\(Uniform)
		\end{tabular}			& --				& --				& --				& 0.1141			& 0.1237 \\
		\hline
		\begin{tabular}[c]{@{}c@{}}
			w.o. defocus
		\end{tabular}			& 0.0522			& 0.0371			& 0.0779			& 0.1061			& 0.1186 \\
		\hline
		\begin{tabular}[c]{@{}c@{}}
			w.o. expsr\\(Gaussian)
		\end{tabular} 			& 0.0537			& 0.0520			& 0.0789			& 0.1173			& 0.1336 \\
		\hline
		\begin{tabular}[c]{@{}c@{}}
			w.o. expsr\\(Uniform)
		\end{tabular} 			& --				& --				& --				& 0.1219			& 0.1355 \\
		\hline
		\begin{tabular}[c]{@{}c@{}}
			NVDiffRecMC
		\end{tabular}  			& 0.0548			& 0.0526			& 0.0835			& \textbf{0.1052}	& \textbf{0.1171} \\
		\hline
		\multicolumn{6}{l}{The bold number marks the best value among all conditions.} \\
	\end{tabular}
	\label{tab:LPIPS_cmpr}
\end{table}

\section{Simulation of AGV in Virtual Environment}
\label{sec:AGV_demo}
As an end-to-end demonstration, we used the reconstructed digital twin of Real Scene 1 in the multi-physics simulator Chrono to simulate an AGV navigating a designated waypoint-based path through the scene. The digital twin comprised the object meshes and their material textures (roughness, metallic, and normals), which had been optimized with the camera simulator (without the defocus blur layer) and were imported into Chrono without modification. The ground mesh, created and isolated as described in Sec.~\ref{subsec:real_scene_setup}, was rendered with a uniform earthy yellow base color and modeled as non-flat but rigid terrain. This mesh can optionally be converted into deformable terrain when loaded into Chrono, in which case Chrono computes terrain deformation based on vehicle-soil terramechanics interaction.

The AGV was a 1:24 down-scaled model of the High Mobility Multipurpose Wheeled Vehicle (HMMWV), including rigid tires \cite{chronoVehicle2019}. The vehicle was tested on a winding path designed to navigate around obstacles. A PID controller was used to regulate both throttle and steering for path-following behavior.

A pinhole camera from Chrono::Sensor was positioned behind the AGV to simulate the perspective of a drone following the vehicle. This camera generated radiance images, which were then processed by the PyTorch-based camera simulator using various camera settings to synthesize photorealistic images. These synthetic images included a range of optical effects, simulating the generation of visual data by a real camera in a dynamic environment. In addition, a third-person camera was placed above the scene to provide a global view, recording the AGV's navigation and interaction with the environment. This simulation demonstrated how the pipeline shown in Fig.~\ref{fig:concat_NeRF} offers a viable solution to the robotics question posed at the beginning of the paper: \textit{How can we build a digital twin of a real off-road environment to evaluate an AGV navigation policy in simulation?}

\begin{figure*}[!t]
	\centering
	\includegraphics[width=1.0\columnwidth]{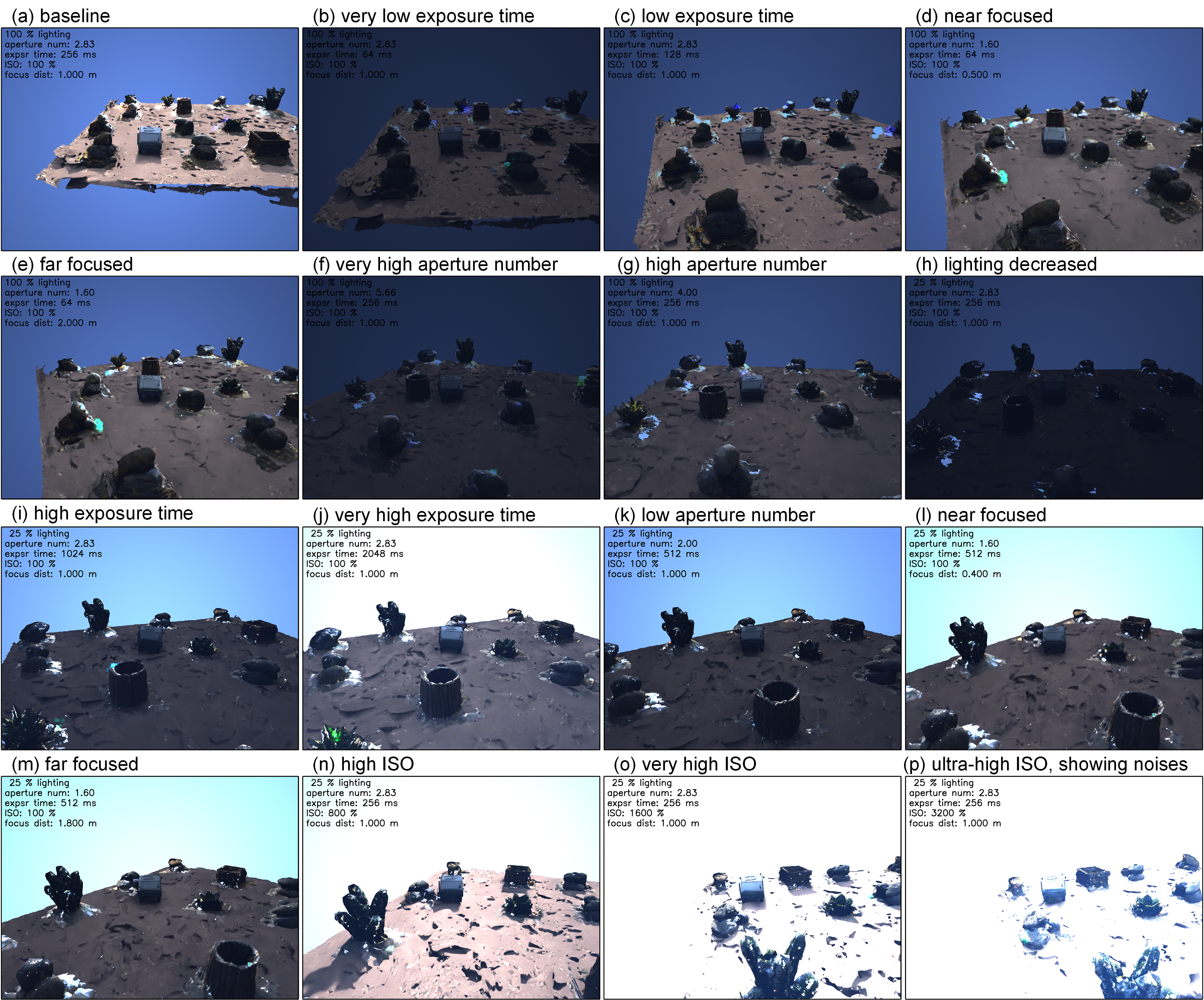}
	\caption{The sequence of synthetic images generated by the camera simulator using various camera settings during the AGV navigating simulation.}
	\label{fig:sim_exp_demo}
\end{figure*}

The attached video, along with the sequence of images in Fig.~\ref{fig:sim_exp_demo} (frames (a) to (p), shown in chronological order), presents the simulation results. With different camera settings applied, Fig.~\ref{fig:sim_exp_demo} demonstrates a wide range of realistic visual effects, including variations in the exposure triangle (aperture, exposure time, ISO), defocus blur, and sensor noise, but excluding motion blur.

However, some color tone bias can be observed in the synthetic images in Fig.~\ref{fig:sim_exp_demo} compared with those in Figs.~\ref{fig:real01_defocus_cmpr} and \ref{fig:real01_expsr_cmpr}. This discrepancy arises since the radiance in Figs.~\ref{fig:real01_defocus_cmpr} and \ref{fig:real01_expsr_cmpr} was generated using NVDiffRecMC, which is the differentiable renderer used for optimization, whereas Fig.~\ref{fig:sim_exp_demo} used Chrono::Sensor. Although both renderers implement BRDFs based on similar physical principles and use identical material parameters (albedo, roughness, and metallic), subtle differences in their rendering algorithms and shading methods lead to noticeable chromatic mismatches.

This outcome highlights a key challenge in renderer transferability: despite NVDiffRecMC's claim that optimized materials and geometry are portable across renderers, discrepancies in underlying shading techniques can introduce significant color bias. Addressing such inconsistencies remains a direction for future research.

\section{Conclusion and future work}
\label{sec:conclusion}
This paper introduced DiffPhysCam, a differentiable, physics-based camera simulator implemented in PyTorch. It exposes a complete set of camera setting parameters (aperture number, exposure time, ISO, focal length, and focus distance) together with tunable model parameters calibrated to a real camera through physical experiments. By providing realistic, parameterized camera simulation, DiffPhysCam supports inverse rendering for novel image synthesis and 3D reconstruction, enabling more accurate and transferable simulation for robotics.

The camera model takes scene radiance as input and sequentially applies lens distortion, vignetting, defocus blur, an aggregator, noise addition, and a camera response function, as well as motion blur. The model parameters were calibrated against a reference camera through physical experiments, after which the synthesized images closely matched real photographs.

We then concatenated the calibrated simulator after the differentiable renderer NVDiffRecMC to optimize mesh geometry and material textures across three simulated and two real-world scenes. Ablation studies showed that the defocus blur component reproduces realistic depth-of-field effects in qualitative comparisons but has limited impact on quantitative metrics, whereas the exposure-related components consistently improve performance across all quantitative evaluations.

Finally, we demonstrated an end-to-end multi-physics simulation combining vehicle dynamics, terramechanics, PID-based control, and photorealistic rendering. A scaled-down HMMWV followed a winding path through the digital twin of the real-world scene while a calibrated virtual camera tracked it and captured realistic images under various settings. 

Several challenges remain to be addressed. First, the current camera simulator models only geometric optics and omits effects governed by physical optics. Extending the simulator to include additional optical artifacts, e.g., chromatic aberration and lens flare, within the existing framework represents a direction for future work. Second, a noticeable performance gap persists between simulation and real-world experiments. Addressing this gap will require further improvements to both the optimization pipeline and the fidelity of the camera simulator. Lastly, the reconstructed 3D scene rendered by Chrono::Sensor exhibits color tone discrepancies compared to those rendered by NVDiffRecMC, despite sharing material and lighting parameters. Bridging this gap between different renderers remains an open research problem.

\section*{Acknowledgments}
\label{sec:acknowledge}
This work was supported in part by the National Science Foundation under grant number CMMI2153855 and by NASA under project STTR-80NSSC24CA030.

The authors thank Tony Adriansen for his substantial assistance in constructing the Real Scene 1 experimental setup and capturing photographs, and Alexis Ruiz for assisting with the Real Scene 2 experiments.

\section*{Appendix}
\appendix
\section{Additional Calibration Results}
\label{sec:appndx_gamma}
\begin{figure}[!t]
	\centering
	\includegraphics[width=0.7\columnwidth]{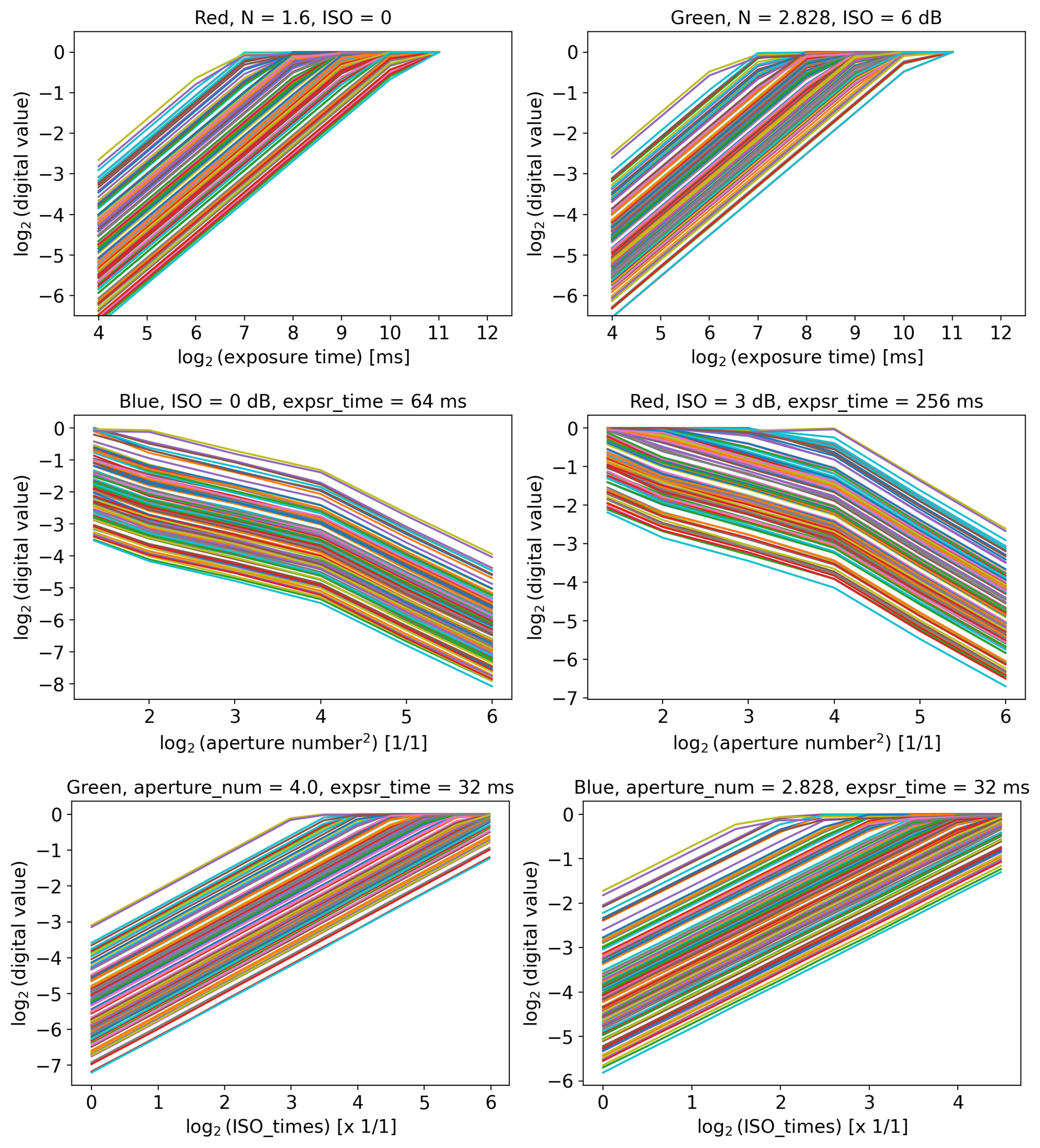}
	\caption{Examples of logarithmic plots showing the relationships between RGB digital values and exposure parameters: exposure time (top row), squared aperture number (middle row), and ISO (bottom row).}
	\label{fig:gamma_expsr_examples}
\end{figure}

This appendix provides the detailed results supporting the gamma calibration of the camera response function summarized in Sec.~\ref{subsubsec:CRF_cali}. The same 216 photos used in Sec.~\ref{subsubsec:exposure_cali}, which were captured under various exposure times, aperture numbers, and ISOs, were employed to calibrate the gamma value in Eq.~\ref{eq:log_dv_to_expsr}. Each photo contained 360 averaged RGB digital values (DVs) for all color blocks as probes. By fixing two of the three exposure control parameters and changing the third one to double or halve the exposure (which is commonly called one \textit{stop}), the logarithmic relationship between DVs and exposure variation can be observed as described in Eq.~\ref{eq:log_dv_to_expsr}. Figure~\ref{fig:gamma_expsr_examples} shows examples of such logarithmic plots, where each unit increment along the horizontal axis corresponds to a doubling of exposure. The plots are very linear, except in those exposure-saturated regions. Note that the plots of the aperture numbers (middle row) are less consistent, since the aperture size was adjusted by hand and thus not sufficiently precise.

\begin{figure}[!t]
	\centering
	\includegraphics[width=0.7\columnwidth]{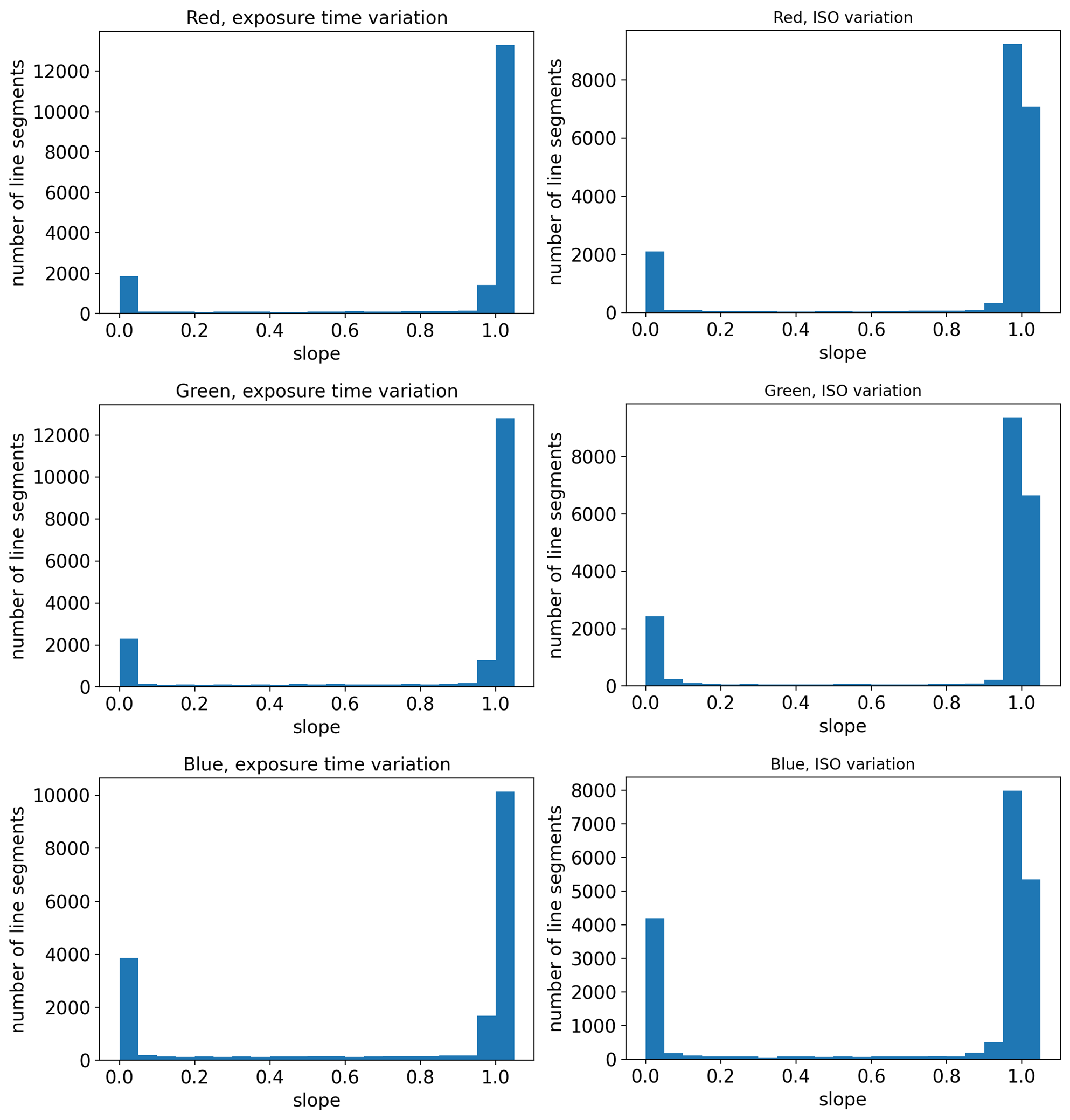}
	\caption{Histograms of slope distributions for RGB channels under exposure time variations (left column) and ISO variations (right column).}
	\label{fig:gamma_hists}
\end{figure}

To determine the gamma value, the slopes of the line segments from plots such as those in Fig.~\ref{fig:gamma_expsr_examples} were collected and are visualized as histograms in Fig.~\ref{fig:gamma_hists}. Aperture-related variations were excluded due to their imprecision. In each histogram, the first bin was also ignored, as it contained saturation outliers. The two most frequent bins were selected to compute the averaged slope, which is the gamma value. The averaged slopes of the three RGB channels for exposure time variation were 1.0080, 1.0065, and 1.0072; for ISO variation, they were 0.9985, 0.9983, and 0.9972, respectively. Rounding all these averaged slopes to the first decimal place yields an overall average slope of 1.0, implying a linear relationship between DVs and exposure amount: $y = ax + b$.

\section{Hyperparameter Setting for Inverse Rendering}
\label{sec:appndx_opt}
This appendix lists the hyperparameter settings used for the inverse rendering optimization described in Sec.~\ref{subsec:metrics}. Vertex positions were not optimized in the second pass. The texture resolution was set to 2048 $\times$ 2048 pixels, and the output image resolution was 768 $\times$ 1024 pixels. The initial tetrahedral grid resolution was 256. The mesh scale was set slightly larger than the spatial range of the scene to ensure that the initial mesh grid fully surrounded the scene. The parameter \textit{n\_samples} was set to 12, and the batch size was 1. The initial learning rates were 0.0250 for Pass~1 and 0.003 for Pass~2 in all scenes except Real Scene 2, which used 0.0150 and 0.0018, respectively. All the other hyperparameters remained at their default values in NVDiffRecMC.


\section*{Data availability}
The authors have shared the links to the data and code in Abstract in the paper.

\section*{Declaration of generative AI and AI-assisted technologies in the writing process}
During the preparation of this work, the authors used ChatGPT for proofreading the language and improving the clarity and composition of the manuscript. The authors reviewed and edited the output as needed and take full responsibility for the content of the published article. 

\bibliographystyle{cas-model2-names}

\bibliography{BibFiles/refsSensors,BibFiles/refsChronoSpecific,BibFiles/refsRobotics,BibFiles/refsSBELspecific,BibFiles/refsML-AI,BibFiles/refsMBS,BibFiles/refsCompSci,BibFiles/refsTerramech,BibFiles/refsFSI,BibFiles/refsDEM,BibFiles/refsGraphics,BibFiles/refsAutonomousVehicles}


\vskip3pc

\end{document}